\documentclass[12pt]{article}
\usepackage{amsmath}
\usepackage{amssymb}
\begin{document}
%\begin{flushright}
%BHU-PHYS-CAS Preprint\\
%arXiv: 1005.5067 [hep-th]
%\end{flushright}
\vskip 2cm
\begin{center}
{\bf {\Large   {Universal Superspace Unitary Operator and Nilpotent 
(Anti-)dual BRST Symmetries: Superfield Formalism}}}

\vskip 3.0cm

{\sf T. Bhanja$^{(a)}$, N. Srinivas$^{(a)}$, R. P. Malik$^{(a,b)}$}\\
$^{(a)}$ {\it Physics Department, Centre of Advanced Studies,}\\
{\it Banaras Hindu University, Varanasi - 221 005, (U.P.), India}\\

\vskip 0.1cm

%\vskip 0.1cm

$^{(b)}$ {\it DST Centre for Interdisciplinary Mathematical Sciences,}\\
{\it Institute of Science, Banaras Hindu University, Varanasi - 221 005, India}\\
{\small {\sf {e-mails: tapobroto.bhanja@gmail.com; seenunamani@gmail.com; rpmalik1995@gmail.com}}}
\end{center}

\vskip 2cm
\noindent
{\bf Abstract:} We exploit the key concepts of  the augmented version of superfield approach to 
Becchi-Rouet-Stora-Tyutin (BRST) formalism to derive the superspace (SUSP) {\it dual} unitary operator (and its 
Hermitian conjugate) and demonstrate their utility in the derivation of the
nilpotent and absolutely anticommuting (anti-)dual BRST symmetry transformations for 
a set of interesting models of the {\it Abelian} 1-form gauge theories.
These models are the one (0+1)-dimensional (1$D$) rigid rotor, modified versions of the two 
(1+1)-dimensional (2$D$) Proca  as well as anomalous gauge theories and 2$D$ model 
of a self-dual bosonic field theory. We show the {\it universality} of the SUSP {\it dual} unitary operator and its 
Hermitian conjugate in the cases of {\it all} the Abelian models under consideration. 
These SUSP dual unitary operators, besides maintaining the explicit group structure,
provide the alternatives to the dual horizontality condition (DHC) and dual gauge invariant restrictions 
(DGIRs) of the superfield formalism. The derivation of the {\it dual} unitary operators
and corresponding (anti-)dual BRST symmetries are completely {\it novel} results in our present investigation.

\vskip 2cm

\noindent
PACS numbers: 11.30.Pb; 03.65.-w; 11.30.-j\\

\noindent
{\it {Keywords}}: Superspace {\it dual} unitary operator; augmented version of superfield formalism;
nilpotency and absolute anticommutativity properties; (anti-)dual BRST symmetry transformations;
geometrical interpretations; 1D and 2D Abelian 1-form gauge theories

\newpage
\section{Introduction}

A {\it classical} gauge theory is endowed with the local gauge symmetries which are generated 
by the {\it first-class} constraints  in the terminology of Dirac's
prescription for the classification scheme [1,2]. Thus, one of the decisive features of a classical
gauge theory is the existence of the first-class constraints on it.
The above cited {\it classical} local gauge symmetries are traded with the {\it quantum} gauge [i.e. (anti-)BRST] 
symmetries within the framework of Becchi-Rouet-Stora-Tyutin (BRST) formalism.  
The existence of the Curci-Ferrari (CF) condition(s) [3] is one of the key signatures 
of a {\it quantum} gauge theory when it is BRST quantized.  
The geometrical superfield approach [4-10] to  BRST formalism is one 
of the most elegant methods which leads to the derivation of the nilpotent
and absolutely anticommuting (anti-)BRST  transformations for a given $D$-dimensional
gauge theory. In addition, this usual superfield formalism
[6-8] also leads to the deduction of the (anti-)BRST invariant CF-conditions
(which are the key signature of the {\it quantum} gauge theories). Thus, we observe that, 
in one stroke, the {\it usual} superfield formalism (USF) produces the 
CF-type condition(s) as well as the proper quantum gauge [i.e., (anti-)BRST] symmetries
for a quantum gauge theory. It is, therefore,
clear that the USF sheds light on various aspects of quantum gauge theories when they are discussed 
within the framework of BRST formalism.

The USF [4-10], however, leads to the 
derivation of nilpotent (anti-)BRST 
symmetry transformations {\it only} for the gauge field and associated (anti-)ghost fields 
of a given {\it quantum} gauge theory. It does not shed any light on the derivation of the  
(anti-)BRST symmetry transformations, associated with the {\it matter} fields, in a given 
{\it interacting} quantum gauge theory where there is a coupling between the gauge field and matter fields. 
In a set of papers [11-15], the above superfield formalism has been {\it consistently} 
extended so as to derive precisely the (anti-)BRST symmetry transfromations for the gauge, 
matter and (anti-)ghost fields {\it together}. Whereas the {\it usual} superfield 
formalism exploits the theoretical potential and power of the horizontality condition (HC),
its extended version utilizes the theoretical strength
of the HC as well as the gauge invariant 
restrictions (GIRs) {\it together} in  a consistent manner. The extended 
version of the USF has been christened [11-15] as the augmented 
version of superfield formalism (AVSF). One of the key observations of the applications of USF and AVSF
is the fact that the group structure of the (non-)Abelian gauge theories remains somewhat {\it hidden}
but the geometry of these theories becomes  quite explicit as we take the help of the cohomological
operators of differential geometry.

The purpose of our present investigation is to exploit the theoretical strength of 
AVSF to derive the superspace {\it dual} unitary operators for the 1$D$ and 2$D$ interesting models
 of the Abelian 1-form gauge theories corresponding to the (anti-)dual BRST [i.e., (anti-)co-BRST]
 symmetry transformations
 which have been shown to exist for the above models. These models are the 1$D$ rigid rotor, modified 
versions of the 2$D$ Proca as well as anomalous gauge theories and 2$D$ self-dual bosonic field theory.  
In fact, these models have been shown to provide the physical examples of Hodge theory within the framework of 
BRST formalsm where the (anti-)BRST as well as (anti-)co-BRST symmetries exist together with a unique bosonic
symmetry and the ghost-scale symmetry [16-20]. The universal superspace (SUSP) {\it unitary} operators, 
corresponding to the nilpotent (anti-)BRST
symmetry transformations, have already been shown to exist for the above models (see, e.g. [21] for details).
The central theme of our present investigation is to derive the SUSP {\it dual} unitary 
operators (from the above universal {\it unitary} operators). 
The derivation of the unitary SUSP operators is important because the group structure of the theory is maintained
and it remains {\it explicit} throughout the whole discussion within the framework of AVSF.
The forms of this  SUSP unitary operators were {\it first} suggested in an earlier work on the
superfield approach to the non-Abelian 1-form gauge theory [7]. These expressions,
however, were intuitively chosen but not derived theoretically. Moreover, the Hermitian 
conjugate unitary operator, corresponding to the chosen SUSP unitary operator, 
was derived after imposing some {\it outside} conditions on the fields and Grassmannian variables
(of the SUSP unitary operator).

In our present investigation, we have derived the {\it dual}  SUSP unitary operators 
(i.e. SUSP dual unitary operator and its Hermitian conjugate) which provide the alternatives to
 the  dual horizontality condition (DHC) and dual gauge invariant restrictions (DGIRs). 
This derivation is a completely {\it new} result because it leads to the derivation of the 
nilpotent and absolutely anticommutating (anti-)co-BRST symmetry transformations which have 
been derived earlier within the framework of superfield approach where the DHC and DGIRs 
have played some decisive roles [22-25]. In fact, we have obtained the proper dual SUSP  
unitary operator (and its Hermitian conjugate) from the {\it universal} unitary operators that 
have been derived in our earlier works [26,27] on {\it interacting} gauge theories. To be 
specific, we have already derived the explicit form of the  SUSP unitary operator 
(and its Hermitian conjugate) in the 4$D$ interacting Abelian 1-form gauge theory with Dirac 
and complex scalar fields [26] as well as 4$D$
non-Abelian gauge theory with Dirac fields [27]. In our present investigation, we have obtained the {\it dual}
 SUSP unitary operator (and its Hermitian conjugate) from the {\it duality} operation on the {\it universal}
SUSP unitary operators (that have already been derived in our earlier work [26] for the 
4$D$ interacting Abelian theory).
The form of the SUSP dual unitary operator (and its Hermitian conjugate) turns out to be {\it universal}
for {\it all} the Abelian 1-form gauge models under consideration which are defined on the 
{\it one} and {\it two} dimensional Minkowskian flat spacetime manifold.

Our present investigation is essential on the following key considerations. First and foremost,
as we have shown the {\it universality} of the SUSP  {\it unitary} operator (and its Hermitian 
conjugate) in the context of the models under consideration for the derivations of the (anti-)BRST
symmetries, similarly, we have to derive the {\it universal} SUSP  {\it dual} unitary operator 
(and its Hermitian conjugate) for the (anti-)co-BRST transformations for the sake of completeness. 
We have accomplished this goal in present investigation. Second, the existence of the  SUSP {\it dual} 
unitary operator (and its Hermitian conjugate) provides the alternatives to the DHC and DGIRs 
that have been invoked in the derivation of the nilpotent (anti-)co-BRST symmetry transformations 
within in the framework of AVSF. One of the highlights of our present investigation
is the observation that the  SUSP {\it dual} unitary operator and its Hermitian conjugate turn 
out to be {\it universal} for {\it all} the {\it Abelian} models that have been considered in 
our present endeavor. Third, the Abelian 1-form theories (that have been considered here) are 
{\it intresting} because these have been shown to provide the physical examples of Hodge theory. 
Fourth, we have found out the (anti-)co-BRST symmetry transformation for a {\it new} model which 
has {\it not} been considered in our earlier works on the superfield approach to BRST formalism 
[22-25]. We have obtained,  for the first time, the (anti-)co-BRST symmetry transformations for 
the modified version of the 2$D$ anomalous gauge theory. Thus, it is a {\it novel} result in our 
present endeavor. Finally, our present attempt is our modest first-step towards our central goal 
of establishing that these  SUSP {\it dual} unitary operators are {\it universal} even in the 
case of {\it non-Abelian} theories.

The contents of our present investigation are organized as follows. In Sec. 2, we briefly discuss the
(anti-)dual BRST symmetry transformations in the Lagrangian fromulation for the 1$D$ rigid rotor, modified
versions of the 2$D$ Proca as well as anomalous gauge theories and 2$D$ self-dual bosonic field theory.
We exploit the theoretical strength of the DHC and DGIRs to derive the above nilpotent symmetries 
within the framework of superfield formalism in Sec. 3. Our
Sec. 4 deals with the derivation of the above nilpotent symmetries by using the SUSP {\it dual} unitary operators. 
In Sec. 5, we summarize our key results and point out a few future directions for further investigation.\\

\noindent
{\it General Notations and Convention}: We adopt the notation $s_{(a)d}$ for the on-shell as well as 
off-shell nilpotent (anti-)dual-BRST [i.e. (anti-)co-BRST] symmetry transformations for {\it all} the
1$D$ and 2$D$ models under consideration. In the description of 2$D$ theories, we choose 2$D$ flat Minkowski
metric with the signature (+1, -1) so that the dot product between two non-null vectors $P_{\mu}$ and $Q_{\mu}$
is defined as: $P \cdot Q = \eta_{\mu\nu}P^{\mu}Q^{\nu}=P_{0}Q_{0}-P_{i}Q_{i}$ where the Greek indices 
$\mu,\nu,\lambda.....=0,1$ correspond to the 2$D$ spacetime directions and the Latin indices $i,j,k....=1$
stand for the space direction {\it only}. Our choice of the Levi-Civita tensor $\varepsilon_{\mu\nu}$ is
such that $\varepsilon_{01}=+1=\varepsilon^{10}$ and $\varepsilon_{\mu\nu}\varepsilon^{\mu\nu}\,=\,+\,2\,!$,
$\varepsilon_{\mu\nu}\varepsilon^{\nu\lambda}\,=\,\delta_{\mu}^{\lambda}$,
$\varepsilon_{\mu\nu}\varepsilon^{\mu\lambda}\,=\,-\,\delta_{\nu}^{\lambda}$, etc. The notations for 
the scalar and superscalar fields have been chosen to be $\phi(x)$ and $\Phi(x,\theta,\bar{\theta})$ 
on the 2$D$ Minkowskian spacetime manifold and (2, 2)-dimensional supermanifold, respectively, for 
all the Abelian models under consideration (wherever these fields are required for discussions).

\section{Preliminaries: (Anti-)dual-BRST Symmetries}

To begin with, we discuss here the nilpotent $(s_{(a)d}^2 = 0)$ and absolutely 
anticommuting $(s_{d}s_{ad}+s_{ad}s_{d} = 0)$ (anti-)dual BRST symmetries $s_{(a)d}$ in the 
Lagrangian formulation for  the 1$D$ rigid rotor which is described by the following first-order 
Lagrangian (see, e.g.[16])
\begin{eqnarray}
L_B = \dot r \, p_r + \dot\vartheta \, p_{\vartheta} - \frac{p_{\vartheta}^2}
{2 r^2} - \lambda \,(r - a) + B\,(\dot \lambda - p_r)
+ \frac{B^2}{2} - i\, \dot{\bar C}\,\dot C + i\, \bar C\, C,
\end{eqnarray}
where the pair $(\dot r, \dot{\vartheta})$ is the generalized velocities corresponding to the
generalized polar coordinates $(r,\vartheta)$ of the rigid rotor. We have taken the unit mass $(m = 1)$
while defining the pair $(p_{r}, p_{\vartheta})$ as the conjugate momenta corresponding to the
coordinates $(r, \vartheta)$. Here $\lambda(t)$ is the ``gauge" variable of the theory
(which is a 1-form ${\lambda}^{(1)} = dt\;\lambda (t)$ on a 1$D$ manifold) and
$B(t)$ is the Nakanishi-Lautrup type auxiliary variable. The anticommuting 
$(C(t)\,\bar{C}(t)+\bar{C}(t)\,C(t) = 0)$ fermionic $(C^2 = \bar{C}^2 = 0)$ (anti-)ghost variables
$(\bar{C})C$ are required to maintain the unitarity in the theory. All the variables are 
the function of an evolution parameter $t$ and an overdot 
($\dot r, \dot{\vartheta}, \dot {\lambda}, \dot C, \dot{\bar{C}},$ etc.) corresponds to a 
single derivative (i.e.\,$\dot r = dr/dt,\, \dot v = dv/dt,$\,etc.) with respect to $t$. It can be readily
checked [16] that, under the following (anti-)dual BRST symmetry transformations $(s_{(a)d})$
\begin{eqnarray}
&&s_{ad}\,\lambda = C, \quad s_{ad}\,p_{r}\,=\,\dot C, \quad s_{ad}\,\bar C = -\;i\;(r-a),\quad
s_{ad}\,[r, \vartheta, B, p_{\vartheta}, C]\,=\,0, \\ \nonumber
&&s_{d}\,\lambda = \bar C, \quad s_{d}\,p_{r}\,=\,\dot{\bar C}, \quad\quad s_{d}\,C = i\;(r-a),\quad \quad
s_{d}\,[r, \vartheta, B, p_{\vartheta}, \bar C]\,=\,0, 
\end{eqnarray}
the Lagrangian (1) and gauge-fixing term remain invariant 
$(s_{(a)d} L_B = 0, s_{(a)d} (\dot\lambda - p_r) = 0)$.

We now focus our attention  on the (anti-)dual BRST symmetry transformations for the modified version of
2$D$ Proca theory (with mass parameter $m$) which is described by the following (anti-)BRST 
invariant Lagrangian density (see, e.g. [17, 18] for details)
\begin{eqnarray}
\mathcal L_{B}^{(p)} \,&=&\,\frac {1}{2} (E\,-\,m\,\tilde \phi)^2\,+\,m\,E\,\tilde\phi\,-\,
\frac {1}{2}\,\partial_{\mu}\,\tilde \phi\,\partial^{\mu}\,\tilde \phi\,+\,\frac {m^2}{2}\,A_{\mu}\,A^{\mu}\,
+\,\frac{1}{2}\,\partial_{\mu}\,\phi\,\partial^{\mu}\,\phi \\ \nonumber
&-&\,m\,A_{\mu}\,\partial^{\mu}\,\phi\,-\,\frac{1}{2}\,(\partial \cdot A\,+\,m\,\phi)^2
\,-\,i\,\partial_{\mu}\,\bar C\,\partial^{\mu}\,C\,+\,i\,m^2\,\bar C\,C,
\end{eqnarray}
where the 1-form $A^{(1)}\,=\,dx^{\mu}\,A_{\mu}$ defines the 2$D$ gauge potential $A_{\mu}$ and the
corresponding curvature tensor $F_{\mu\nu}$ is defined from the 2-form $F^{(2)}\,=\,dA^{(1)}\,=\,
[(dx^{\mu}\wedge dx^{\nu}/2!)\,F_{\mu\nu}]$ where $d\,=\,dx^{\mu}\partial_{\mu}$ (with $d^2\,=\,0$)
is the exterior derivative. In 2$D$, the curvature tensor $F_{\mu\nu}$ contains only {\it one} 
independent component which is nothing but the electric field $E$. The latter turns out to
be a {\it pseudoscalar} in two (1 + 1)-dimensions of spacetime. In the above, we have a pair
$(\phi, \tilde \phi)$ of fields which is constructed by a scalar Stueckelberg field $\phi$ and
a pseudoscalar field $\tilde \phi$. The latter has been introduced in the theory on the physical as well as 
mathematical grounds [17,18]. The fermionic $(C^2\,=\,\bar C^2\,=\,0, \,\,C\,\bar C\,+\,\bar C\,C\,=\,0)$
fields are the (anti-)ghost fields $(\bar C)\,C$ which are required to maintain the unitarity
in the theory. 
It can be readily checked that the following nilpotent ($ s_{(a)d}^2 = 0$) and absolutely anticommuting
$(s_{d} s_{ad}\,+\,s_{ad} s_{d}\,=\,0)$ (anti-)dual BRST symmetry transformations $(s_{(a)d})$
\begin{eqnarray}
&&s_{ad}\,A_{\mu}\,=\,-\varepsilon_{\mu\nu}\,\partial^\nu\,C, \quad s_{ad}\,C\,=\,0,\quad s_{ad}\,\bar C
\,=\,i\,(E\,-\,m\,\tilde \phi), \\ \nonumber
&&s_{ad}\,E\,=\,\square\, C,\quad s_{ad}\,(\partial \cdot A\,+\,m\,\phi)\,=\,0,\quad 
s_{ad}\,\phi\,=\,0, \quad s_{ad}\,\tilde \phi\,=\,-m\,C, \\\nonumber
&&s_{d}\,A_{\mu}\,=\,-\varepsilon_{\mu\nu}\,\partial^\nu\,\bar C, \quad s_{d}\,\bar C\,=\,0,
\quad s_{d}\,C\,=\,-\,i\,(E\,-\,m\,\tilde \phi), \\ \nonumber
&&s_{d}\,E\,=\,\square\, \bar C,\quad s_{d}\,(\partial \cdot A\,+\,m\,\phi)\,=\,0,\quad 
s_{d}\,\phi\,=\,0, \quad s_{d}\,\tilde \phi\,=\,-m\,\bar C,
\end{eqnarray}
leave the action integral invariant because the  Lagrangian density transforms
 to the  total spacetime derivatives (see, e.g. [17],[18] for details).
It is to be noted that the total gauge-fixing term remains invariant under $s_{(a)d}$
[i.e. $\,s_{(a)d}\,(\partial \cdot A + m \phi) = 0$].

Another modified version of the 2$D$ Abelian 1-form model is the bosonized version of
anomalous Abelian 1-form gauge theory which is described by the following (anti-)BRST 
invariant Lagrangian density (see, e.g., [19] for details)
\begin{eqnarray}
\mathcal{L}_{B}^{(a)}\,&=&\,-\,\frac{1}{4}\,F^{\mu\nu}\,F_{\mu\nu}\,
+\,\frac{1}{2}\,\partial_{\mu}\phi\,\partial^{\mu}\phi\,
+\,\frac{a}{2}\,A_{\mu}\,A^{\mu}\\ \nonumber
&+&\,(\eta^{\mu\nu}\,-\,\varepsilon^{\mu\nu})\,\partial_{\mu}\phi\,A_{\nu}\,+\,
\sigma \,[(a-1)\,(\partial \cdot A)\,+\,\varepsilon^{\mu\nu}\,\partial_{\mu} \,A_{\nu}] \\ \nonumber
&+&\frac{(a-1)}{2}\,\partial_{\mu}\,\sigma\,\partial^{\mu}\,\sigma\,+\,B\,(\partial \cdot A)\,+\,
\frac{B^2}{2} \,+\,i\,\partial_{\mu}\,\bar C\,\partial^{\mu}\,C,
\end{eqnarray}
where, as explained earlier, the 2-form curvature $F_{\mu\nu}$ has only electric field as its
existing component and $a$ is the ambiguity parameter in the regularization of the fermionic determinant
when the 2$D$ chiral Schwinger model (with electric charge $e = 1$) is bosonized in terms of the scalar
field $\phi$. We have introduced an extra 2$D$ bosonic field $\sigma(x)$ in the theory to convert
the second-class constraints of the original 2$D$ chiral Schwinger model into the first-class system
so that we could have the ``classical" gauge and ``quantum" (anti-)BRST symmetries 
in the theory (see, e.g. [19] for details).
The other symbols $(\bar C)C$ and $B(x)$ have already been explained earlier. 
The Lagrangian density (5) can be re-expressed as
\begin{eqnarray}
\mathcal{L}_{B}^{(a)}\,&=&\,{\cal B} E\,-\,\frac{{\cal B}^2}{2}
+\,\frac{1}{2}\,\partial_{\mu}\phi\,\partial^{\mu}\phi\,
+\,\frac{a}{2}\,A_{\mu}\,A^{\mu}\\ \nonumber
&+&\,(\eta^{\mu\nu}\,-\,\varepsilon^{\mu\nu})\,\partial_{\mu}\phi\,A_{\nu}\,+\,
\sigma \,\Bigl[(a-1)\,(\partial \cdot A)\,+\,\varepsilon^{\mu\nu}\,\partial_{\mu}\,A_{\nu}\Bigr] \\ \nonumber
&+&\frac{(a-1)}{2}\,\partial_{\mu}\,\sigma\,\partial^{\mu}\,\sigma\,+\,B\,(\partial \cdot A)\,+\,\frac{B^2}{2}
\,+\,i\,\partial_{\mu}\,\bar C\,\partial^{\mu}\,C,
\end{eqnarray}
which is endowed with the following (anti-)co-BRST symmetries
\begin{eqnarray}
&&s_{ad}\,A_{\mu}\,=\,-\varepsilon_{\mu\nu}\,\partial^{\nu}\,C, \quad s_{ad}C\,=\,0, 
\quad s_{ad}\bar C\,=\,-i{\cal B}, \quad s_{ad}{\cal B }\,=\,0,\quad s_{ad}\phi\,=\,- C, \\ \nonumber
&&s_{ad}E\,=\,\square C,\quad s_{ad}(\partial \cdot A)\,=\,0, s_{ad}B\,=\,0, \quad
s_{ad}\sigma\,=\,-\frac{C}{a-1} \cong C \; (1 + a),\\ \nonumber
&&s_{d}A_{\mu}\,=\,-\varepsilon_{\mu\nu}\partial^{\nu} {\bar C}, \quad s_{d}\bar C\,=\,0,
\quad s_{d}C\,=\,i{\cal B}, \quad s_{d}{\cal B}\,=\,0, \quad s_{d}\phi\,=\,-\bar C, \\ \nonumber
&& s_{d}E\,=\,\square \bar C, \quad s_{d}(\partial \cdot A)\,=\,0,\quad s_{d}B\,=\,0, \quad
s_{d}\sigma\,=\,-\frac{\bar C}{(a-1)} \, \cong \bar C \; (1 + a),
\end{eqnarray}
where we have introduced an auxiliary field ${\cal B}(x)$ to linearize the kinetic term 
($-\,\frac{1}{4}\,F_{\mu\nu}F^{\mu\nu} = \frac{1}{2}\,E^{2} \equiv {\cal B}\,E
- \frac{1}{2}\,{\cal B}^2$) of our modified 2$D$ anomalous Abelian 1-form gauge theory.
The symmetry invariance can be explicitly checked, by using the above transformations, where 
the action integral $S=\int d^{2}x\,{\cal L}_{B}^{(a)}$ remains invariant because the above
Lagrangian density transforms to the total spacetime derivatives (see, e.g. [19] for details).

Finally, we concentrate on a theoretically interesting system of the Abelian 1-form model of 
the 2$D$ self-dual bosonic field theory which is described by the following (anti-)BRST 
invariant Lagrangian density (see, e.g. [20] for details)
\begin{eqnarray}
\mathcal {L}_{B}^{(s)}\,&=&\,\frac{1}{2}\,\dot{\phi}^{2}\,-\,\frac{1}{2}\,
\dot{v}^{2}\,+\,\dot{v}\,(v'\,-\,\phi')\,+\,\lambda\,[\dot{\phi}\,-\,\dot{v}\,
+\, v'\,-\,\phi']\nonumber\\
& - &\,\frac{1}{2}\,(\phi'\,-\,v')^{2} - \, \frac{1}{2}\,(\dot{\lambda}\,-\,v\,-\,\phi)^2\,
-\,i\,\dot{\bar{C}}\,\dot{C}\,+\,2\,i\,\bar{C}\,C,
\end{eqnarray}
where an overdot on fields (e.g. 
$\dot v\,=\,\partial v/\partial t,\,\dot\phi\,=\,\partial\phi/\partial t$)
corresponds to the expression for the ``generalized" velocities  (where a derivative with 
respect to the evolution  parameter $t$ is taken into account) and the prime on the fields
($\phi'\,=\,\partial\phi/\partial x,\,v'\,=\,\partial v/\partial x$)
is the space derivative with respect to the space coordinate $x$. Here $v(x)$ field
is the Wess-Zumino (WZ) field and $\phi(x)$ field is the 2$D$ self-dual bosonic field (of our present
2$D$ self-dual field theory). Rest of the symbols have already been explained earlier. The above
Lagrangian density (8) is endowed with the following (anti-)dual-BRST symmetry transformations ($s_{(a)d}$)
\begin{eqnarray}
&&s_{ad}\,\lambda\,=\,C,\qquad s_{ad}\,\phi\,=\,\frac{\dot C}{2},\qquad s_{ad}\,v\,=\,\frac{\dot C}{2},
\qquad \quad s_{ad}\,C\,=\,0,\\\nonumber
&&s_{ad}\,\bar C\,=\,\frac{i}{2}\,(\dot \phi\,-\,\dot v\,+\,v'\,-\,\phi'),\quad
s_{ad}\,(\dot\phi\,-\,\dot v\,+\,v'\,-\,\phi')\,=\,0, \\ \nonumber
&&s_{d}\,\lambda\,=\,\bar C, \qquad \quad s_{d}\,\phi\,=\,\frac{\dot{\bar C}} {2}, \qquad \quad
s_{d}\,v\,=\,\frac{\dot {\bar C}}{2}, \quad \qquad s_{d}\,\bar C\,=\,0,\\ \nonumber
&&s_{d}\,C\,=\,-\frac{i}{2}(\dot\phi\,-\,\dot v\,+\,v'\,-\,\phi'),\qquad
s_{d}\,(\dot \phi\,-\,\dot v\,+\,v'\,-\,\phi')\,=\,0,
\end{eqnarray}
because the Lagrangian density (8) transforms to the total ``time" derivatives as 
\begin{eqnarray}
&&s_{ad}\,{\cal L}_{B}^{(s)}\,=\,\frac{\partial}{\partial t}\,\Big[\frac {\dot C}{2}\,
(\dot \phi\,-\,\dot v\,+\,v'\,-\,\phi')\Big],\\ \nonumber
&&s_{d}\,{\cal L}_{B}^{(s)}\,=\,\frac{\partial}{\partial t}\,\Big[\frac {\dot {\bar C}}{2}\,
(\dot \phi\,-\,\dot v\,+\,v'\,-\,\phi')\Big].
\end{eqnarray}
Thus, the action integral $S\,=\,\int d^{2}x\,{\cal L}_{B}^{(s)}$ remains invariant under $s_{(a)d}$
for the physical fields that vanish off at $t = \pm \infty $.

The decisive features of the (anti-)dual BRST [i.e. (anti-)co-BRST] symmetry transformations
are the observations that (i) they are nilpotent of order two (i.e. $ s_{(a)d}^2\,=\,0$) which demonstrates
their fermionic nature, 
(ii) these nilpotent symmetries are also absolutely anticommuting 
$(s_{d}\,s_{ad}\,+\,s_{ad}\,s_{d}\,=\,0)$ in nature that shows the linear independent
of $s_{d}$ and $s_{ad}$, (iii) the gauge-fixing terms, owing their origin to the co-exterior derivative 
(see, e.g. [16-20]), remain invariant under the (anti-)dual BRST 
symmetry transformations $(s_{(a)d})$. Thus, the nomenclature (anti-)co-BRST symmetries
is appropriate for these symmetries. This observation should be contrasted with the (anti-)BRST
symmetries where the total kinetic term, owing its origin to the exterior derivative, remains invariant [16-20].

\section{Nilpotent (Anti-)co-BRST Symmetries: Superfield Approach to the Abelian 1-Form Gauge Theories}

We briefly discuss here the derivation of the (anti-)co-BRST symmetries of our 1-form gauge theories
by exploiting the geometrical superfield approach to BRST formalism [4-15]. First of all, we 
focus on the derivation of the above symmetries in the context of 1$D$ rigid rotor.
In this connection, we note that the gauge-fixing term (${\dot\lambda}-p_{r}$) remains invariant
under $s_{(a)d}$. Furthermore, we observe that this term has its geometrical origin in the 
co-exterior derivative ($\delta$) because
$\delta \lambda^{(1)}\,=\,\ast \,d \ast (dt \lambda(t)) \equiv {\dot \lambda(t)}$ 
where $(\ast )$ is the Hodge duality operation on the 1$D$ manifold. Here we have taken the 1-form as: 
$\lambda^{(1)}=dt\lambda(t)$. According to the basic tenets of AVSF, the invariance of the gauge-fixing
term implies that this quantity should remain independent of the ``soul" coordinates ($\theta, \bar\theta$)
when we generalize it onto the (1, 2)-dimensional supermanifold parameterized by the 
superspace coordinates ($t, \theta, \bar\theta$) where the pair ($\theta, \bar\theta$) 
is a set of Grassmannian variables (with ${\theta}^2 =  {\bar\theta}^2 = 0,\,
\theta\bar\theta + \bar\theta\theta = 0$). In older literature [28], the latter
coordinates have been christened as the ``soul" coordinates and $t$ has been called as 
the body coordinate. In other words, we have the following equality
\begin{eqnarray}
\star \,{\tilde d} \star   \tilde \lambda^{(1)}(t,\theta, \bar{\theta})\,-\,P_{r}(t,\theta,\bar{\theta})\,=\,
\ast  \,d \ast  \lambda^{(1)}(t)\,-\,p_{r}(t),
\end{eqnarray}
where $\star  $ is the Hodge duality operation on the (1, 2)-dimensional supermanifold on which our 1$D$
ordinary theory is generalized. The other quantities in the equation (11) are 
\begin{eqnarray}
&&d\,=\,dt\,\partial_{t} \longrightarrow \tilde d \,=\,dt\,\partial_{t}\,+\,d\theta\,\partial_{\theta}\,+\,
d\bar\theta\,\partial_{\bar\theta},\qquad 
p_{r}(t)\longrightarrow P_{r}(t,\theta,\bar\theta)\\ \nonumber
&&\lambda^{(1)}\,=\,dt\,\lambda(t)\longrightarrow \tilde \lambda^{(1)}\,=\,dt\,\Lambda(t,\theta,\bar\theta)\,+\,
d\theta\,{\bar F}(t,\theta,\bar\theta)\,+\,d{\bar\theta}\,F(t,\theta,\bar\theta),
\end{eqnarray}
where the supervariables $\Lambda(t,\theta,\bar\theta),\, F(t,\theta,\bar\theta),\, \bar F(t,\theta,\bar\theta)$
and $P_{r}(t,\theta,\bar\theta)$ have the following expansions along the $(\theta, \bar\theta)$-directions of
(1, 2)-dimensional supermanifold [25]:
\begin{eqnarray}
&&\Lambda(t,\theta,\bar\theta)\,=\,\lambda(t)\,+\,\theta {\bar R(t)}\,+\,{\bar\theta}R(t)\,
+\,i\,\theta\,{\bar\theta}S(t),\\ \nonumber
&&F(t,\theta,\bar\theta)\,=\,C(t)\,+\,i\,\theta\,{\bar B}_{1}(t)\,
+\,i\,\bar\theta B_{1}(t)\,+\,i\,\theta \,\bar\theta \,s(t), \\\nonumber
&&\bar F(t,\theta,\bar\theta)\,=\,\bar C(t)\,+\,i\,\theta\,{\bar B}_{2}(t)\,
+\,i\,\bar\theta B_{2}(t)\,+\,i\,\theta \, \bar\theta \,\bar s(t), \\\nonumber
&&P_{r}(t,\theta,\bar\theta)\,=\,p_{r}(t)\,+\,\theta\,{\bar K}(t),
+\,\bar\theta K(t)\,+\,i\,\theta \,\bar\theta \, L(t).
\end{eqnarray}
We note, in the above, that the secondary variables ($R, \bar R, s, \bar s, K, \bar K$) are fermionic and 
($S, B_{1}, {\bar B}_{1}, B_{2}, {\bar B}_{2}, L$) are bosonic in nature. It is elementary to verify that,
in the limit $\theta\,=\,\bar\theta = 0$, we get back our 1D variables ($\lambda, C, \bar C, p_{r}$)
that are present in the Lagrangian (1). The dual horizontality condition (DHC) [cf. (11)]
leads to the following [25]
\begin{eqnarray}
{\bar B}_{1}\,=\,B_{2}\,=\,0,\quad s\,=\,\bar s = 0, \quad
B_{1}\,+\,{\bar B}_{2}\,=\,0,\quad {\bar K}\,=\,\dot{\bar R},\quad K = \dot R,\quad L = \dot S.
\end{eqnarray} 
The above relationships prove that some of the secondary variables are zero and others are interconnected
in a definite and precise manner. It is worthwhile to mention that the condition 
$B_{1}\,+\,{\bar B}_{2}\,=\,0$ is the {\it trivial} CF-type condition. This restriction is a 
{\it physical} condition in our theory because it is an (anti-)co-BRST invariant quantity.

We resort to the additional restrictions on the supervariables that are motivated by the basic requirements
of AVSF which state that the (anti-)co-BRST invariant quantities should be independent of 
the ``soul" coordinates. In this connection, we observe the following
\begin{eqnarray} 
s_{(a)d}\,\bigl[ \dot r\,p_r - i\, \dot{\bar C}\, \dot C\bigr] = 0, \qquad\qquad s_{(a)d}\,\bigl[ \lambda\,(r - a)  
- i\, \bar{C}\, C\bigr] = 0,
\end{eqnarray}
which, ultimately, imply the following equalities  due to DGIRs, namely;
\begin{eqnarray}
&&\dot{R} (t, \theta, \bar\theta) \, P^{(R)}_{r} (t, \theta, \bar\theta)
-i\,\dot{\bar{F}}^{(R)} (t, \theta, \bar\theta)\, 
\dot{F}^{(R)} (t, \theta, \bar\theta) = \dot{r}\,p_{r} - i\,\dot{\bar{C}}\,\dot{C},\\ \nonumber
&&\Lambda (t, \theta, \bar\theta) \,[R (t, \theta, \bar\theta) - a] 
- i\,\bar{F}^{(R)} (t, \theta, \bar\theta) \,F^{(R)} (t, \theta, \bar\theta)  = \,\lambda\,(r - a) - i\, \bar{C}\,C,
\end{eqnarray}
where the {\it new} notations (with $R (t, \theta, \bar\theta) = r( t)$) are explicitly written as
\begin{eqnarray} 
&& F^{(R)}(t, \theta, \bar{\theta}) = C(t) + i\,\bar\theta\, {\cal B},\qquad
{\bar {F}}^{(R)}(t, \theta, \bar{\theta}) = \bar C(t) - i\,\theta\, {\cal B},\nonumber\\
&& P^{(R)} (t, \theta, \bar{\theta}) = p_{r}(t) + \theta \,(\dot{\bar R}) 
+ \bar\theta\, (\dot R) + i\,\theta\,\bar\theta (\dot S). 
\end{eqnarray}
In the above, we have chosen $B_1(t) = -\,\bar B_2(t) = -\,{\cal B}$ and taken the
inputs from (14). The conditions (15) are now supported by
the observations: $s_{d}\,(\lambda \bar C) = 0$ and $s_{ad}\,(\lambda C) = 0$. These two conditions 
lead to the following restrictions on the supervariables:
\begin{eqnarray}
\Lambda (t, \theta, \bar\theta)\,{\bar {F}}^{(R)}(t, \theta, \bar{\theta}) = 
\lambda (t)\, \bar C (t), 
\qquad \Lambda (t, \theta, \bar\theta)\,F^{(R)}(t, \theta, \bar{\theta}) = 
\lambda (t)\, C (t).
\end{eqnarray}
Finally, we obtain the expressions for the secondary variables in terms of
the original variables of the Lagrangian (1) as (see, e.g. [25] for details):
\begin{equation}
\bar{R} = C,\quad\qquad R = \bar{C},\qquad\quad {\cal B} = (r - a),\qquad\quad S = (r - a)\,\equiv {\cal B}.
\end{equation}
The substitution of these values into the expansions (13) and (17) leads to the following 
{\it final} expressions for the expansion of the supervariables (see, e.g.  [25] for details)
\begin{eqnarray}
\Lambda^{(d)}\,(t, \theta, \bar{\theta}) &=& \lambda(t) + \theta\, (C) + \bar{\theta}\, (\bar{C}) 
 +  \theta\, \bar{\theta}\,
[i\, (r - a)] \nonumber\\
&\equiv & \lambda(t) + \theta\, (s_{ad}\, \lambda) + \bar{\theta}\,(s_{d}\, \lambda) 
+ \theta\,\bar{\theta}\, (s_{d}\,s_{ad}\,\lambda), \nonumber\\
F^{(d)}\,(t, \theta, \bar{\theta}) &=& C(t) + \theta\, (0) + \bar{\theta}\, [i\,(r - a)]  
+  \theta\, \bar{\theta}\, (0) \nonumber\\
&\equiv & C(t) + \theta\, (s_{ad}\, C) + \bar{\theta}\,(s_{d}\, C) 
+ \theta\,\bar{\theta}\, (s_{d}\,s_{ad}\,C), \nonumber\\
{\bar F}^{(d)}\,(t, \theta, \bar{\theta}) &=& \bar{C}(t) + \theta\, [-\,i\,(r - a)] 
+ \bar{\theta}\,(0)  +  \theta\, \bar{\theta}\, (0) \nonumber\\
&\equiv & \bar{C}(t) + \theta\, (s_{ad}\, \bar{C}) + \bar{\theta}\,(s_{d}\, \bar{C}) 
+ \theta\,\bar{\theta}\, (s_{d}\,s_{ad}\,\bar{C}),\nonumber\\
P^{(d)}_{r}(t, \theta, \bar{\theta}) &=& p_{r}(t) + \theta\, (\dot{C}) 
+ \bar{\theta}\, (\dot{\bar{C}}) 
+ \theta\,\bar{\theta}\, (i \dot r)\nonumber\\
& \equiv & p_{r}(t) + \theta\,(s_{ad}\,p_{r}) + \bar{\theta}\, (s_{d}\,p_{r}) 
+ \theta\,\bar{\theta}\,(s_{d}\,s_{ad}\,p_{r}),
\end{eqnarray}
where the superscript $(d)$ on the supervariables denotes the expansions that have
been obtained after the application of DHC and DGIRs. A careful and close look at the above
expansions demonstrate that we have already obtained the non-trival (anti-)co-BRST symmetry
transformations for the variables ($\lambda, C, \bar C, p_{r}$) of the 1$D$ rigid rotor. The trivial
nilpotent (anti-)co-BRST symmetry transformations $s_{(a)d} [r, p_{\theta}, \theta] = 0$ are self-evident.
It is clear that there is a geometrical meaning of $s_{(a)d}$ in the language of translational
operators ($\partial_\theta, \partial_{\bar\theta}$) along the Grassmannian directions 
($\theta, \bar\theta$) of the (1, 2)-dimensional supermanifold. The nilpotency 
(${\partial_\theta}^2 = {\partial_{\bar\theta}}^2 = 0$) and absolute anticommutativity
($\partial_\theta \,\partial_{\bar\theta} +\partial_{\bar\theta}\,\partial_\theta$)
of these generators provide the geometrical meaning to the nilpotency ($s_{(a)d}^2 = 0$)
and absolute anticommutativity ($s_d\;s_{ad} + s_{ad}\;s_d = 0$) of the (anti-)co-BRST
symmetries.

We now focus on the derivation of the (anti-)co-BRST symmetry transformations ($s_{(a)d}$) in the context of
the modified versions of the 2$D$ Proca and anomalous
Abelian 1-form gauge theories within the framework of AVSF. 
In this connection, first of all, we observe that the gauge-fixing term ($\partial \cdot A\,\pm \,m\phi$)
remains invariant [i.e. $s_{(a)d}\,(\partial \cdot A \pm m \phi)\,=\,0$] under $s_{(a)d}$ (because, separately
and independently, we have: $s_{(a)d}(\partial \cdot A)\,=\,0, \, s_{(a)d}\phi\,=\,0$). We note that 
$(\partial \cdot A)$ has its origin in the co-exterior derivative ($\delta$) because 
$\delta A^{(1)} = -\,\ast d \ast (dx^{\mu}\,A_{\mu})\,=\,(\partial \cdot A)$. Thus, we have to generalize
this relationship onto the (2, 2)-dimensional supermanifold parametrized by the superspace co-ordinates 
($x^{\mu}, \theta, \bar\theta$). Thus, according to the basic tenets of AVSF, we have the following equality
(see, e.g. [18] for details)
\begin{eqnarray}
\star \,{\tilde d} \star   \tilde A^{(1)}=\,\ast  \,d \ast  A^{(1)},
\qquad \quad\phi(x) \to \Phi(x, \theta, \bar\theta)\,=\,\phi(x),
\end{eqnarray}
where $\star$ is the Hodge duality operation on the (2, 2)-dimensional supermanifold and other 
relevant symbols have already been explained earlier. In our earlier works [18], the l. h. s. of relation
(21) has been already computed clearly by taking the help of the Hodge duality operation $\star$
defined on the (2, 2)-dimensional supermanifold [22].

At this stage, we would like to clarify some of the {\it new} symbols used in equation (21). 
We have the generalization of the ordinary exterior derivative $d = dx^{\mu}\,\partial_{\mu}$ and
Abelian 1-form $A^{(1)} = dx^{\mu}\,A_{\mu}$ onto the (2, 2)-dimensional supermanifold as
\begin{eqnarray}
&&d\,=\,dx^{\mu}\,\partial_{\mu}\rightarrow\,\tilde{d}\,=\,dx^{\mu}\,\partial_{\mu}\,+
\,d\theta\,\partial_{\theta}\,+\,d\bar{\theta}\,\partial_{\bar{\theta}},\nonumber\\
&&A^{(1)}\,=\,dx^{\mu}\,A_{\mu}\,\rightarrow\,\tilde{A}^{(1)}\,
=\,dx^{\mu}\,{\cal B}_{\mu}(x,\, \theta, \bar{\theta})\,+\,d\theta\,\bar{F}(x, \theta, \bar{\theta})\,
+\,d\bar{\theta}\,F(x, \theta, \bar{\theta}),
\end{eqnarray}
where the superfields ${\cal B}_{\mu}(x, \theta, \bar\theta), F(x, \theta, \bar\theta)$ and 
$\bar F (x, \theta, \bar\theta)$ have the following expansions along ($\theta, \bar\theta$)-directions 
of the (2, 2)-dimensional supermanifold 
\begin{eqnarray}
&&{\cal B}_{\mu}(x, \theta , \bar{\theta})\, = \,A_{\mu}(x)\,+
\,\theta\,R_{\mu}^{(1)}(x)\,+\,\bar{\theta}\,R_{\mu}^{(2)}(x)\,
+\,i\,\theta\,\bar{\theta}\,S_{\mu}(x),\nonumber\\
&&F(x, \theta , \bar{\theta})\, = \,C(x)\,+\,i\,\theta\, B_{1}(x)\,+\,i\,\bar{\theta}\,B_{2}(x) \, 
+\,i\,\theta\,\bar{\theta}\,s(x),\nonumber\\
&&\bar{F}(x, \theta , \bar{\theta})\,= \,\bar{C}(x)\,+\,i\,\theta\, B_{3}(x)\,+
\,i\,\bar{\theta}\,B_{4}(x)+\,i\,\theta\,\bar{\theta}\,\bar s(x),
\end{eqnarray}
where ($A_{\mu}(x), C(x), \bar C(x)$) are the basic fields of the modified versions of the 2$D$
{\it Proca} and {\it anomalous gauge theories}. The set of secondary fields 
($R_{\mu}^{(1)}, R_{\mu}^{(2)}, s, \bar s$) are fermionic and ($S_{\mu}, B_{1}, B_{2}, B_{3}, B_{4}$)
are bosonic in nature (because of the fermionic nature of the Grassmannian variable ($\theta, \bar\theta$)).
The dual horizontality condition (21) leads to the following very useful relationships
 (see, e.g. [18] for details)
\begin{eqnarray}
&&\partial \cdot R^{(1)}\,=\,\partial \cdot R^{(2)}\,=\,\partial \cdot S \,=\,0, \quad 
s\,=\,\bar s\,=\,0, \\ \nonumber
&&B_{1}\,=\,B_{4}\,=\,0,\quad B_{2}\,+\,B_{3}\,=\,0,
\end{eqnarray}
where the relation $B_{2}\,+\,B_{3}\,=\,0$ is like the CF-type condition which turns out to be a 
{\it trivial} relationship. We would like to state that the details of the equation (24) have been
worked out in our earlier work on the superfield approach to the modified version of 2$D$ Proca theory
[18]. The interesting point is that the above conditions are true in the AVSF approach to the modified
version of 2$D$ anomalous gauge theory, too.

The above relations do {\it not} lead to the exact form of $R_{\mu}^{(1)}, R_{\mu}^{(2)}, S_{\mu}$ and 
($B_{2}, B_{3}$). The CF-type condition $B_{2}\,+\,B_{3}\,=\,0$ allows us to choose $B_{2}\,=\,\cal B$ 
so that $B_{3}\,=\,-\,\cal B$. Now, we exploit the virtue of the AVSF  to derive the
{\it exact} forms of the secondary fields and observe that the following (anti-)co-BRST invariant quantity
\begin{eqnarray}
s_{(a)d}\,\Big[\varepsilon^{\mu\nu}\,(\partial_{\mu}{\cal B)}\,A_{\nu}\,
-\,i\,\partial_{\mu}\,\bar C \partial^{\mu} C\Big]\,=\,0,
\end{eqnarray}
permits us to demand that the superfield generalization of the above quantity on the (2, 2)-dimensional
supermanifold must be independent of the soul co-ordinates ($\theta, \bar\theta$). Thus, we have the 
following equality
\begin{eqnarray}
\varepsilon^{\mu\nu}\,\bigl (\partial_{\mu}{\cal B} (x) \bigr )\,{\cal B}_{\nu}(x, \theta, \bar\theta)\,
-\,i\,\partial_{\mu}\,{\bar F}^{(d)}(x, \theta, \bar\theta)\, 
\partial^{\mu} F^{(d)}(x, \theta, \bar\theta)\\\nonumber 
\equiv \varepsilon^{\mu\nu}\,(\partial_{\mu}{\cal B} (x)) \,A_{\nu}(x)\,
-\,i\,\partial_{\mu}\,\bar C(x)\, \partial^{\mu} C(x).
\end{eqnarray}
In the above, the expansions for the superfields $F^{(d)} (x, \theta, \bar\theta)$  and
$\bar F^{(d)} (x, \theta, \bar\theta)$ are
\begin{eqnarray}
&&F^{(d)} (x, \theta, \bar\theta)\,=\,C(x)\,+\,\bar\theta\,(-\,i\,{\cal B}) \equiv C(x) + 
\bar\theta\, (s_d \, C), \nonumber\\
&&{\bar F}^{(d)} (x, \theta, \bar\theta)\,=\,\bar C(x)\,+\,\theta\,(i\,{\cal B})
\equiv \bar C(x) + \theta\, (s_{ad}\, \bar C),
\end{eqnarray}
because it is clear from (24) that $s = \bar s = 0$ and $B_{1} = B_{4} = 0$.
The substitution of the explicit expansion of $F^{(d)}(x, \theta, \bar\theta), {\bar F}^{(d)}(x, \theta, \bar\theta)$
and ${\cal B}_{\mu}(x, \theta, \bar\theta)$ into (26) leads to the following relationships when we 
equate the coefficients of $\theta, \bar\theta$ and $\theta \bar\theta$ equal to zero, namely;
\begin{eqnarray}
&&\varepsilon^{\mu\nu}\,\bigl(\partial_{\mu}{\cal B}(x)\bigr)\,\bar R_{\nu}(x)\,
+\partial_{\mu}C(x) \,\partial^{\mu} {\cal B}(x)\,=\,0, \\ \nonumber
&&\varepsilon^{\mu\nu}\,\bigl(\partial_{\mu}{\cal B}(x)\bigr)\, R_{\nu}(x)\,
+\partial_{\mu}\bar C(x) \,\partial^{\mu} {\cal B}(x)\,=\,0, \\ \nonumber
&&\varepsilon^{\mu\nu}\,\bigl(\partial_{\mu}{\cal B}(x)\bigr)\,S_{\nu}(x)\,
+\partial_{\mu}{\cal B}(x) \,\partial^{\mu} {\cal B}(x)\,=\,0,
\end{eqnarray}
leading to the final determination of the secondary fields (with the help from (24)) as 
\begin{eqnarray}
{\bar R}_{\mu}(x)\,=\,-\,\varepsilon_{\mu\nu}\,\partial^{\nu}C(x), \quad
R_{\mu}(x)\,=\,-\,\varepsilon_{\mu\nu}\,\partial^{\nu}\,{\bar C}(x), \quad
 S_{\mu}(x)\,=\,-\,\varepsilon_{\mu\nu}\,\partial^{\nu}{\cal B}(x). 
\end{eqnarray}
Thus, we have the following explicit expansions of the superfield:
\begin{eqnarray}
{\cal B}_{\mu}^{(d)}(x, \theta, \bar\theta)\,&=&\,A_{\mu}(x)\,+\,\theta\,
\bigl(-\,\varepsilon_{\mu\nu}\,\partial^{\nu}C(x)\bigr)\,+\,\bar \theta \,
\bigl(-\,\varepsilon_{\mu\nu}\,\partial^{\nu}{\bar C}(x)\bigr)
\,+\,\theta\bar\theta\,[-\,i \,\varepsilon_{\mu\nu}\,\partial^{\nu}{\cal B}(x)]\nonumber\\
&\equiv& \,A_{\mu}(x)\,+\,\theta\,
\bigl(s_{ad}\,A_{\mu}(x)\bigr)\,+\,\bar \theta \,
\bigl(s_{d}\,A_{\mu}(x)\bigr) 
\,+\,\theta\bar\theta\,\bigl(s_{d}\,s_{ad}\, A_{\mu}(x)\bigr) .
\end{eqnarray}
It is very clear that we have derived the following (anti-)co-BRST symmetry transformations
for the fields ($A_\mu (x), C (x), \bar C (x)$) due to superfield formalism:
\begin{eqnarray}
&&s_{d}\,A_{\mu}\,=\,-\,\varepsilon_{\mu\nu}\,\partial^{\nu}\bar C,\quad 
s_{ad}\,A_{\mu}\,=\,-\,\varepsilon_{\mu\nu}\,\partial^{\nu} C,
\quad s_{d}s_{ad}\,A_{\mu}\,=\,-\,i\,\varepsilon_{\mu\nu}\,\partial^{\nu}\cal B, \nonumber \\
&&s_{d}\,C\,=\,-\,i\,{\cal B}, \qquad s_{ad}\,C\,=\,0, \qquad s_{d}s_{ad}\,C\,=\,0, 
\qquad \\ \nonumber 
&& s_{d}\,\bar C = 0, 
\qquad s_{ad}\,\bar C = \,i\,{\cal B}, \qquad s_{d}s_{ad}\,\bar C = 0.
\end{eqnarray}
The nilpotency and absolute anticommutativity properties of the (anti-)co-BRST symmetry 
transformations imply that we have $s_{(a)d}\,{\cal B} = 0$. Up to this point, our results are 
{\it same} for the superfield description of the modified versions of 2D 
{\it Proca} and {\it anomalous gauge theories} because the above transformations are common
for both the theories.

To determine the (anti-)dual-BRST symmetry transformations for the $\tilde \phi(x)$ field of
the modified version of the 2$D$ Proca theory (cf. Eq. (3)), we observe
that $s_{(a)d}\,\bigl[A_{\mu} - \frac {1}{m}\,\varepsilon_{\mu\nu}\,\partial^{\nu}{\tilde \phi}\bigr] = 0$.
Thus, according to the basic requirements of AVSF, we demand that this quantity should remain
independent of the ``soul" coordinates, namely;
\begin{eqnarray}
{\cal B}_{\mu}^{(d)}(x, \theta, \bar\theta) - \frac {1}{m}\,\varepsilon_{\mu\nu}\,\partial^{\nu}{\tilde \Phi}
(x, \theta, \bar\theta) = 
A_{\mu} (x) - \frac {1}{m}\,\varepsilon_{\mu\nu}\,\partial^{\nu}{\tilde \phi}(x).
\end{eqnarray}
Now if we taken the expansion of the superfield
\begin{eqnarray}
{\tilde \Phi}(x, \theta, \bar\theta) = \tilde \phi(x) + \theta\,f_{4}(x)\,
+\,\bar\theta\,f_{5}(x)\,+\,i\,\theta\,\bar\theta\,b_{4}(x),
\end{eqnarray}
we obtain, from (32), the relationships 
$f_{4}(x) = -\,m\,C, \, f_{5}(x) = -\,m\,\bar C, \, b_{4}(x) = -\,m\,{\cal B}(x)$
which show the fermionic nature of $(f_4, f_5)$ and bosonic nature of $b_4$.
Thus, the final expansion of (33), in terms of the (anti-)co-BRST symmetries $s_{(a)d}$, is
\begin{eqnarray}
{\tilde \Phi}^{(d)}(x, \theta, \bar\theta) & = & \tilde \phi(x) + \theta\,(-\,m C)\,
+\,\bar\theta\,(-\,m \bar C)\,+\,\theta\,\bar\theta\,(i\,m {\cal B})\\ \nonumber
& \equiv & \tilde \phi(x) + \theta\,(s_{ad}\tilde \phi)\,
+\,\bar\theta\,(s_{d}\tilde \phi)\,+\,\theta\,\bar\theta\,(s_{d}\,s_{ad}\tilde \phi).
\end{eqnarray}
We have, therefore, derived {\it all} the non-trival (anti-)co-BRST symmetry transformations for the fields
$A_{\mu}, C, \bar C$ and $\tilde \phi$ of the modified version of 2$D$ Proca theory. The rest
of the transformations are trivial (e.g. $s_{(a)d} \phi = 0$ and $s_{(a)d} {\cal B} = 0$) and
they can be derived in a straightforward manner
from the AVSF because $\phi (x) \to \Phi (x, \theta, \bar\theta) = \phi (x),
{\cal B} (x) \to \tilde {\cal B} (x, \theta, \bar\theta) = {\cal B} (x) $. 
We re-emphasize that the transformations (31) are {\it common} to the
modified versions of 2$D$ Proca and anomalous gauge theories. As far as the latter theory is concerned,
we have to derive the (anti-)co-BRST symmetry transformations for the scalar fields $\phi(x)$ and $\sigma(x)$.
In this connection, we observe that the following useful quantities are (anti-)co-BRST invariant, namely;
$s_{(a)d}\,[E + \square \phi] = 0, \quad s_{(a)d}\,[(a-1)\,\sigma\, - \phi(x)] = 0$.
It is to be noted that 
$E = -\,\varepsilon^{\mu\nu}\,\partial_{\mu} A_{\nu} = \partial_{0}A_{1} - \partial_{1} A_{0}$ in 2$D$.
 Thus, according to the basic requirement of AVSF, we have the following equality due to the restriction
 on the superfield
\begin{eqnarray}
-\,\varepsilon^{\mu\nu}\,\partial_{\mu}{\cal B}_{\nu}^{(d)}(x, \theta, \bar\theta) 
+ \square\,\Phi(x, \theta, \bar\theta) = -\,\varepsilon^{\mu\nu}\,\partial_{\mu} A_{\nu}(x) 
+ \square\,\phi(x),
\end{eqnarray}
where the expansions for ${\cal B}_{\mu}^{(d)}(x, \theta, \bar\theta)$ and $\Phi(x, \theta, \bar\theta)$
are given in (30) and (42) (see below). Substitution of these values into (35) yields the following relationships:
$\bar f_{1} = -\,C, \quad f_{1} = -\,\bar C, \quad b_{1} = -\,{\cal B}$ which imply the following
expansions for the scalar superfield $\Phi^{(d)}(x, \theta, \bar\theta)$ after 
the application of DGIRs, namely;
\begin{eqnarray}
\Phi^{(d)}(x, \theta, \bar\theta) &=& \phi(x) + \theta (-\,C) + \bar \theta (-\,\bar C) 
+ \theta \bar\theta (-i\,{\cal B})\nonumber\\
&\equiv & \phi (x) + \theta\,(s_{ad}\, \phi)\,
+\,\bar\theta\,(s_{d}\, \phi)\,+\,\theta\,\bar\theta\,(s_{d}\,s_{ad}\, \phi).
\end{eqnarray}
We have to determine the (anti-)co-BRST transformations on the field $\sigma (x)$. In this regards,
we have the following equality due to AVSF
\begin{eqnarray}
(a-1)\,\Sigma (x, \theta, \bar\theta) - \Phi^{(d)}(x, \theta, \bar\theta) = (a-1)\,\sigma(x) - \phi(x),
\end{eqnarray}
where the expansion of $\Phi^{(d)}(x, \theta, \bar\theta)$ is given in (36) and we have taken the following
general expansions of $\Sigma (x, \theta, \bar\theta)$ along the Grassmannian
 $(\theta, \bar\theta)$-directions of the (2, 2)-dimensional supermanifold, namely;
\begin{eqnarray}
\Sigma (x, \theta, \bar\theta) = \sigma(x) + \theta\,\bar P(x) + \bar\theta P(x) 
+ i\,\theta\, \bar\theta\, Q(x),
\end{eqnarray}
where the secondary fields $(P(x), \bar P(x))$ are fermionic and $Q$ is bosonic 
(due to the fermionic nature of $\theta$ and $\bar\theta$). It is straightforward to observe,
from (37), that we have:
\begin{eqnarray}
P(x) = \frac{-\,\bar C}{(a-1)}, \qquad \bar P(x) = \frac{-C}{(a-1)},\qquad Q = \frac{-\,{\cal B}}{(a-1)}.
\end{eqnarray} 
The above values imply that the super expansions (38) is
\begin{eqnarray}
\Sigma^{(d)} (x, \theta, \bar\theta) = \sigma(x) + \theta (s_{ad}\,\sigma) + \bar\theta (s_{d}\,\sigma) 
+ \theta \,\bar\theta \, (s_{d}s_{ad}\,\sigma(x)),
\end{eqnarray}
where $s_{d}\,\sigma = -\,(\frac{\bar C}{(a-1)}),\quad s_{ad}\,\sigma = -\,(\frac{C}{(a-1)}), \quad
s_{d}s_{ad} = -\,(\frac{i\,{\cal B}}{(a-1)})$. Thus, we have derived {\it all} the non-trivial 
(anti-)co-BRST symmetry transformations for the modified version of 2$D$ anomalous gauge theory
within the framework of AVSF.

We are now in the position to discuss the superfield approach to the derivation of the (anti-)co-BRST
symmetries for the 2$D$ self-dual chiral bosonic field theory. First of all, we generalize the relevant
fields of the 2$D$ theory onto the (2, 2)-dimensional superfield parametrized by the superspace co-ordinates
($x^{\mu}, \theta, \bar\theta$) as 
\begin{eqnarray}
&&\phi(x)\rightarrow \Phi(x, \theta, \bar\theta), \quad v(x)\rightarrow V(x, \theta, \bar\theta),\quad
C(x)\rightarrow F(x, \theta, \bar\theta) \quad \\ \nonumber
&&\bar C(x)\rightarrow {\bar F}(x, \theta, \bar\theta) \quad
\lambda(x)\rightarrow \Lambda(x, \theta, \bar\theta),
\end{eqnarray}
which have the following expansions along the Grassmannian directions (i.e. ($\theta, \bar\theta$)-directions)
 of the (2, 2)-dimensional supermanifold [20]
\begin{eqnarray}
&& \Phi (x, \theta, \bar\theta) = \phi(x) + i\, \theta \,\bar f_1(x) + i\, \bar\theta\, f_1 (x)
 + i\, \theta\,\bar\theta\, b_1(x), \nonumber\\
&& V (x, \theta, \bar\theta ) = v(x) + i\, \theta \,\bar f_2(x) +i\, \bar\theta\, f_2(x) 
+ i\, \theta\,\bar\theta\, b_2(x), \nonumber\\
&& F(x, \theta, \bar\theta ) = C(x) + i\, \theta\,\bar B_1 (x) + i\, \bar\theta\, B_1(x) 
+ i\, \theta\,\bar \theta\, s(x), \nonumber\\
&& \bar F(x, \theta, \bar\theta ) = \bar C(x) + i\, \theta\, \bar B_2(x) + i\, \bar\theta\, B_2(x) 
+ i\, \theta\,\bar\theta\, \bar s(x), \nonumber\\
&& \Lambda (x, \theta, \bar\theta) = \lambda(x) + \theta\, \bar R(x) + \bar\theta \,R(x) 
+ i\, \theta\, \bar\theta\, S(x),
\end{eqnarray}
where the set ($ S, B_1, \bar B_1, B_2, \bar B_2, b_1, b_2 $) is made up of the bosonic secondary fields and the
fermionic secondary fields are ($ R, \bar R, s,  \bar s , f_1, \bar f_1, f_2, \bar f_2  $).
We obtain the basic fields $(\lambda, \phi, v, C, \bar C)$  of the theory in the
limit  $\theta = \bar\theta = 0$. We shall obtain the exact expressions for the secondary
fields in terms of the basic and auxiliary fields of the theory by exploiting the 
physically motivated restrictions on the superfields. First of all, we take into account the
 appropriate generalizations of the exterior 
derivative and connection 1-form onto the (2, 2)-dimensional supermanifold, as [20]:
\begin{eqnarray}
&&d\longrightarrow \tilde d =   dt\,\partial _ t + d\theta\, \partial _\theta 
+ d\bar\theta\, \partial _{\bar\theta},\nonumber\\
&&\lambda^{(1)} \longrightarrow \tilde \lambda^{(1)} = dt\,\Lambda(x, \theta, \bar\theta) 
+ d\theta\, \bar F(x, \theta, \bar\theta) + d\bar\theta\, F(x, \theta, \bar\theta).
\end{eqnarray}
It should be noted that even though we have generalized the ordinary theory onto the (2, 2)-dimensional
supermanifold, the super exterior derivative ($\tilde d$) has been defined on the (1, 2)-dimensional
super sub-manifold. This is due to the peculiarity of the gauge field in the case of 2$D$ self-dual
bosonic field theory where only one component of the 2$D$ gauge field couples with the matter fields
but the other component of the gauge field remains {\it inert} (see, e.g. [20] for details).
The basic tenets of AVSF state that {\it all} the (anti-)co-BRST invariant quantities
 should be independent of the ``soul" coordinates ($\theta, \bar\theta$). In this 
 connection, we note that the following are the (anti-)co-BRST invariant quantities
 (see, eg. [20] for details)
 \begin{eqnarray}
s_{(a)d}\,\bigl[\phi - v\bigr] = 0,
\qquad s_{(a)d}\,\bigl[\dot\lambda\, - \phi - v\bigr] = 0,  \qquad 
s_{(a)d} \, \bigl[\dot\phi - \dot v + v' - \phi') \bigr] = 0. 
\end{eqnarray}
Thus, the above quantities in the square brackets, when generalized on the (2, 2)-dimensional
supermanifold, should be independent of the ``soul" coordinates ($\theta, \bar\theta$).
Plugging in the expansions from (42), we obtain the following 
\begin{eqnarray}
&&\bar f_1 = \bar f_2\equiv \bar f, \qquad\qquad f_1 = f_2\equiv f, 
\qquad\qquad b_1 = b_2 \equiv b, \nonumber\\
&&\dot{\bar R} = 2\,i\, \bar f, \quad\qquad\qquad \dot{R} = 2\,i\, f, \quad\qquad\qquad \dot S = 2\,b.
\end{eqnarray}
We shall see that these relationships would be useful in our further discussions. For instance, we observe that
the following are the invariant quantities:
\begin{eqnarray}
s_{(a)d}\,\bigl[\dot\lambda\, - 2\,\phi\bigr] = 0, \qquad s_{(a)d}\,\bigl[\dot\lambda\, - 2\, v\bigr] = 0.
\end{eqnarray}
In the above expressions, it is elementary to note that $\delta \,\lambda^{(1)}\,=\,
+\,\ast\,\,d\, \ast \, \lambda^{(1)}$ is nothing but $\dot\lambda$ 
\big(i.e. $ \delta \,\lambda^{(1)}\,=\, +\,\ast\,\,d\, \ast \, (dt\, \lambda (x)) = \dot\lambda (x)$\big). 
We have to generalize this relationship on the (2, 2)-dimensional supermanifold as
\begin{eqnarray}
\tilde\delta\, {\tilde \lambda}^{(1)} (x, \theta, \bar\theta) - 2\,\Phi (x, \theta, \bar\theta) = 
\delta \,\lambda^{(1)} (x) -\,2\, \phi (x), \qquad \quad \lambda^{(1)} (x) = dt\, \lambda (x),
\end{eqnarray}
where $\tilde\delta =\, \star\, \tilde d\, \star $. Here $\star $ is the Hodge duality operation on the  
(1, 2)-dimensional super-submanifold of the general (2, 2)-dimensional supermanifold
and $\tilde\delta $ is the super co-exterior derivative 
(with $\tilde d = dt\,\partial_t\, +\, d\,\theta \,\partial_\theta\, + d\,\bar\theta \,\partial_{\bar\theta} $).
It is to be noted that the gauge field $\lambda$ is a function of $x^\mu (\mu = 0, 1)$ but
the geometrical  quantities $\tilde d$ and $\tilde\delta$ as well as $d = dt\,\partial_t$ and
$\delta = \ast\,d\,\ast$ are defined in terms of $t$ only. In other words, $d$ and $\delta$ are
defined on the 1$D$ sub-manifold of the 2$D$ ordinary Minkowskian spacetime manifold and 
$\tilde d$ and $\tilde\delta$ are defined on the (1, 2)-dimensional super-submanifold of the
(2, 2)-dimensional supermanifold on which our ordinary 2$D$ theory is generalized.
The l.h.s. of (47) has been worked out in our earlier work. The following relationship
emerges from (47):
\begin{eqnarray}
B_1 + \bar B_2 = 0 \qquad  \Longrightarrow \qquad B_1 = - {\cal B} = - \bar B_2. 
\end{eqnarray}
This condition is nothing but the analogue of the CF-type restriction which is essential as
far as the proof of the absolute anticommutativity property (i.e. $s_d s_{ad} + s_{ad} s_d = 0$) of the nilpotent
(anti-)dual BRST symmetry transformations $s_{(a)d}$ is concerned. This condition is also (anti-)dual BRST
invariant under the above  symmetry transformations $s_{(a)d}$ . Thus, this restriction 
is a {\it physical} condition on the
model under consideration within the realm of BRST formalism. In fact, the whole theory is defined
on the constrained hypersurface (defined by the above trivial constrained condition) that is embedded in the
2$D$ Minkowskian spacetime manifold on which the whole of our present theory is defined.

Ultimately, we concentrate on the following (anti-)co-BRST invariance 
\begin{eqnarray}
s_{(a)d}\,\bigl[\lambda(\dot\phi - \dot v + v^\prime - {\phi}^\prime) + 2\,i\,\bar C \, C\bigr] = 0,
\end{eqnarray}
which imply the following restrictions on the supervariables 
\begin{eqnarray}
\Lambda\,\bigl[\dot{\Phi} - \dot{ V} + {V}' -
 {\Phi}' \bigr] + 2\,i\,\bar F^{(d)}\,\bar F^{(d)}
= \lambda\,(\dot\phi - \dot v + v^\prime - {\phi}^\prime) + 2\,i\,\bar C \, C,
\end{eqnarray}
where the expansions for $F^{(d)}$ and $\bar F^{(d)}$ are as follows:
\begin{eqnarray}
&& F^{(d)}(x, \theta, \bar\theta)\, = \,C(x) \, + \, \bar\theta\, (-\,i\,{\cal B}(x)) 
\,\equiv \, C(x)\, + \, \bar\theta\, (s_d\, C(x)),\nonumber\\ 
&& \bar F^{(d)}(x, \theta, \bar\theta) \,= \,\bar C(x) \,+ \,\theta\, (+\,i\,{\cal B}(x)) \, \equiv 
\, \bar C(x)\, + \,\theta \,(s_{ad}\, \bar C(x)).
\end{eqnarray}
Here the superscript $(d)$ denotes the super-expansions obtained after the application 
of DHC given in (47). Plugging in the expressions from (43) and (51), we obtain
\begin{eqnarray}
\bar R = + \, C,\qquad\qquad \quad R = + \,\bar C, \,\qquad\qquad\,\, S = -\,{\cal B},
\end{eqnarray}
which imply the following:
\begin{eqnarray}
 f = - \frac{i}{2}\; \dot {\bar C}, \qquad\quad \bar f = - \frac{i}{2}
\; \dot C,\qquad\quad b = -\,\frac{1}{2}\,\dot{\cal B}.
\end{eqnarray}
At this stage, we are free to choose the auxiliary field ${\cal B}$ in such a manner
that $s_{(a)d} {\cal B} = 0$. The latter condition is essential because of the requirements of 
nilpotency and absolute anticommutativity. We choose the following in terms of the basic fields as
\begin{eqnarray}
{\cal B} = +\, \frac{1}{2}\, \bigl[\dot\phi - \dot v + v^\prime - {\phi}^\prime \bigr],
\end{eqnarray}
which serves our purpose. Finally, we have the following expansions (see, e.g. [20])
\begin{eqnarray}
{\Lambda}^{(d)} (x, \theta, \bar\theta) &=& \lambda(x) + \theta\, (C) + \bar\theta \,(\bar C)
+ \,\theta\, \bar\theta\, \bigl[-\,\frac{i}{2}\,(\dot\phi - \dot v + v^\prime - {\phi}^\prime)\bigr] \nonumber\\
&\equiv&  \lambda(x) + \theta\,(s_{ad}\,\lambda)+ \bar\theta \,(s_{d}\,\lambda) 
+ \theta\, \bar\theta\,\,(s_{d}\,s_{ad}\,\lambda), \nonumber\\
F^{(d)}(x, \theta, \bar\theta ) &=& C(x) + \theta(0) + \bar\theta\, (-\frac{i}{2}\bigl[\dot\phi - \dot v +
 v^\prime - {\phi}^\prime\bigr])
+ \theta\,\bar \theta\, (0) \nonumber\\
&\equiv& C(x) + \theta\,(s_{ad}\,C)+ \bar\theta \,(s_{d}\,C) + \theta\, \bar\theta\,\,(s_{d}\,s_{ad}\,C), \nonumber\\
\bar F^{(d)}(x, \theta, \bar\theta ) &=& \bar C(x) +  \theta\, (\frac{i}{2}\,\bigl[\dot\phi 
- \dot v + v^\prime - {\phi}^\prime\bigr]) + \bar\theta\, (0)
+ \theta\,\bar\theta\,(0) \nonumber\\
&\equiv& \bar C(x) + \theta\,(s_{ad}\,\bar C)+ \bar\theta \,(s_{d}\,\bar C) 
+ \theta\, \bar\theta\,\,(s_{d}\,s_{ad}\,\bar C), \nonumber\\
\Phi^{(d)} (x, \theta, \bar\theta) &=& \phi(x) + \theta \,(+\, \frac{\dot C}{2}) +
 \bar\theta\,(+\, \frac{\dot {\bar C}}{2})
+ \theta\,\bar\theta\,(-\,\frac{i}{4}\,\frac{\partial}{\partial t}\;\bigl[\dot\phi - \dot v +
 v^\prime - {\phi}^\prime\bigr]) \nonumber\\
&\equiv& \phi(x)  + \theta\,(s_{ad}\,\phi)+ \bar\theta \,(s_{d}\,\phi) + 
\theta\, \bar\theta\,\,(s_{d}\,s_{ad}\,\phi), \nonumber\\
V^{(d)} (x, \theta, \bar\theta ) &=& v(x) + \theta \,(+\, \frac{\dot C}{2}) 
+ \bar\theta\,(+\, \frac{\dot{\bar C}}{2})
 + \theta\,\bar\theta\,(-\,\frac{i}{4}\,\frac{\partial}{\partial t}\;\bigl[\dot\phi 
- \dot v + v^\prime - {\phi}^\prime\bigr])\nonumber\\
&\equiv& v(x)  + \theta\,(s_{ad}\,v)+ \bar\theta \,(s_{d}\,v) + \theta\, \bar\theta\,\,(s_{d}\,s_{ad}\,v),
\end{eqnarray}
where the superscript $(d)$ denotes the expansion of the superfields after the imposition
of the DHC and DGIRs within the framework of AVSF.
Thus, we note that we have derived {\it all} the (anti-)co-BRST symmetry transformations listed in Eq. (9).
The nilpotency and absolute anticmmutativity of $s_{(a)d}$ implies that 
$s_{(a)d} \bigl[\dot\phi - \dot v + v' - {\phi}'\bigr] = 0$.

\section{SUSP Dual Unitary Operator: Universal Aspects}

The precise expressions for the SUSP {\it unitary} operator and its Hermitian conjugate have been explicitly
derived in  our earlier work [26] on the 4$D$ interacting {\it Abelian} 1-form gauge theory with Dirac
and complex scalar fields where we have provided the alternatives to the HC and GIRs in the context
of the derivation of the (anti-)BRST symmetries of this theory. These forms are  
expressed, in terms of the familiar symbols, as follows
\begin{eqnarray}
&&U\,(x, \theta, \bar{\theta})\,=\,1+\,\theta\,(-\,i\,\bar{C})\,+\,\bar{\theta}\,(-\,i\,C)\,
+\,\theta\,\bar{\theta}\,(B\,-\, C\,\bar C),\nonumber\\
&&U^{\dagger}\,(x, \theta, \bar{\theta})\,=\,1+\,\theta\,(i\,\bar{C})\,+\,\bar{\theta}\,(i\,C)\,
+\,\theta\,\bar{\theta}\,(-\,B\,-\,C\,\bar {C}),
\end{eqnarray}
which satisfy $UU^{\dagger} = U^{\dagger}U = 1$. It is important to point out that the above
explicit expressions have been derived by exploiting the theoretical strength behind the concept
of {\it covariant derivatives}. The expressions (56) can be also written in the exponential 
forms as
\begin{eqnarray}
&&U\,(x, \theta, \bar{\theta})\,=\,exp\,[\theta\,(-\,i\,\bar{C})\,+
\,\bar{\theta}\,(-\,i\,C)\,+\,\theta\,\bar{\theta}\,B],\nonumber\\
&&U^{\dagger}\,(x, \theta, \bar{\theta})\,=\,exp\,[\theta\,(i\,\bar{C})\,+
\,\bar{\theta}\,(i\,C)\,-\,\theta\,\bar{\theta}\,B],
\end{eqnarray} 
which very clearly demonstrate the validity of unitary condition: $UU^{\dagger} = U^{\dagger}U = 1$.
The basic idea behind the covariant derivative {\it also} leads to the transformation property of the
1-form $A^{(1)} = dx^{\mu} A_{\mu}$ gauge connection under the (anti-)BRST symmetry transformations, 
in the language of SUSP unitary operator and its Hermitian conjugate, as [21,26]
\begin{eqnarray}
\tilde A^{(1)}_{(h)}\,=\,U\,(x, \theta, \bar{\theta})\,A^{(1)}\,(x)\,
U^{\dagger}\,(x, \theta, \bar{\theta})\, +\,i\,\tilde{d}\,U\,(x, \theta, \bar{\theta})
\,U^{\dagger}\,(x, \theta, \bar{\theta}),
\end{eqnarray}
where
${\tilde A}_{(h)}^{(1)} = dx^{\mu}\,{\cal B}_{\mu}^{(h)}(x, \theta, \bar\theta)\,
+\,d\theta\,\bar F^{(h)}(x, \theta, \bar\theta)\,+\,d\bar\theta\, F^{(h)}(x, \theta, \bar\theta)$.
In this expression, the superfield ${\cal B}_{\mu}^{(h)}(x, \theta, \bar\theta)$ yields the (anti-)BRST
symmetry transformations for the gauge field $A_{\mu}(x)$ and the superfields
($F^{(h)}(x, \theta, \bar\theta), \bar F^{(h)}(x, \theta, \bar\theta)$) yield the (anti-)BRST
symmetry transformations for the ghost and anti-ghost fields, respectively. Here the superscript 
$(h)$ denotes the expressions of the superfields after the application of the HC. The equation (58)
provides an alternative to the HC in terms of the SUSP unitary operator $U$ and its Hermitian conjugate
$U^{\dagger}$ (see, e.g. [21] for details). We shall see below that
we can derive the proper (anti-)dual-BRST symmetry transformations for the relevant fields/variables 
from the equations {\it like} (56), (57) and (58) which would be obtained after the application of the
{\it duality} transformations 
($A_{\mu} \to -\,\varepsilon_{\mu\nu}\,A^{\nu},\,C \to \bar C,\,\bar C \to C$).

We focus, first of all, on the derivation of the (anti-)co-BRST symmetry transformations for 
the 1$D$ rigid rotor where the form of the unitary operator and its 
Hermitian conjugate is same as given in (56) and (57) with the replacement 
$x \to t$ (i.e. $U\,(x, \theta, \bar{\theta})|_{x = t} = U\,(t, \theta, \bar{\theta}) $),
where all the fields are functions of $t$ {\it only} (i.e. $B(t), C(t), \bar C (t)$). 
There is a {\it duality} in the theory where $\lambda \to p_{r},\,C \to \bar C$ and 
$\bar C \to C$ for the presence of the (anti-)dual-BRST symmetry transformations 
$s_{(a)d}$. This is due to the fact that the role of $\lambda, p_{r}, C$ and $\bar C$
change in a symmetrical fashion when we go from the (anti-)BRST symmetries to the 
(anti-)dual-BRST symmetries. A careful and close look at Eqn. (2) shows that 
the role of $B$ in the (anti-)BRST symmetry transformations is traded with $(r - a)$ 
in the (anti-) co-BRST symmetries. Thus, we have the SUSP {\it dual} unitary operator 
and its Hermitian  conjugate operator from the unitary operators (56) (with the 
replacement $ B \to (r-a)$) as
\begin{eqnarray}
&&U\,(t, \theta, \bar{\theta})\, \to\, \tilde U\,(t, \theta, \bar{\theta})\,=\,1+\,\theta\,(-\,i\,C)\,+
\,\bar{\theta}\,(-\,i\,\bar C)\,
+\,\theta\,\bar{\theta}\,\bigl[(r - a)\,-\,\bar C\,C\bigl],\nonumber\\
&&U^\dagger \,(t, \theta, \bar{\theta})\,\to\,\tilde U^{\dagger}\,(t, \theta, \bar{\theta})\,=\,1
+\,\theta\,(i\,C)\,+\,\bar{\theta}\,(i\,\bar C)\, +\,\theta\,\bar{\theta}\,\bigl[-\,(r - a)\,-\,\bar{C}\,{C}\bigl],
\end{eqnarray}
which also satisfy ${\tilde U}\,{\tilde U}^{\dagger} = {\tilde U}^{\dagger}\, {\tilde U} = 1$
and they can be exponentiated as 
\begin{eqnarray}
&&\tilde U\,(t, \theta, \bar{\theta})\,=\,exp\,[\theta\,(-\,i\,C)\,+
\,\bar{\theta}\,(-\,i\,\bar C)\,+\,\theta\,\bar{\theta}\,(r - a)],\nonumber\\
&&\tilde U^{\dagger}\,(t, \theta, \bar{\theta})\,=\,exp\,[\theta\,(i\,C)\,+
\,\bar{\theta}\,(i\,\bar C)\,+\,\theta\,\bar{\theta}\,\{-\,(r - a)\}],
\end{eqnarray}
so that we have the validity of unitary condition 
$\tilde U \tilde U^{\dagger} = \tilde U^{\dagger}\tilde U = 1$ in a straightforward manner.
Now the DHC can be expressed in the following fashion
\begin{eqnarray}
p_r^{(1)}(t) \to \tilde P_r^{(1){(d)}}(t, \theta, \bar{\theta})\,=
\,\tilde U\,(t, \theta, \bar{\theta})\,p_r^{(1)}\,(t)\,
\tilde U^{\dagger}\,(t, \theta, \bar{\theta})\, +\,i\,\bigl(\tilde{d}\,\tilde U\,(t, \theta, \bar{\theta})\bigr)
\,\tilde U^{\dagger}\,(t, \theta, \bar{\theta}),
\end{eqnarray}
where $\tilde P_r^{(1){(d)}}(t, \theta, \bar{\theta}) = dt\, P_r^{(d)}(t, \theta, \bar{\theta})\,+
\,d\theta\,F^{(d)}(t, \theta, \bar{\theta})\,+\,d\bar\theta \bar F^{(d)}(t, \theta, \bar{\theta})$. 
It should be noted that we have already taken into account the dual transformations
\begin{eqnarray}
F^{(d)}(t, \theta, \bar{\theta}) \to \bar F^{(d)}(t, \theta, \bar{\theta}),
\qquad\qquad \bar F^{(d)}(t, \theta, \bar{\theta}) \to  F^{(d)}(t, \theta, \bar{\theta}),
\end{eqnarray}
in the definition of the l.h.s. of Eq. (61) which yields the expressions of the
superfields $(P_r^{(d)}, \, F^{(d)},\, \bar F^{(d)})$ after the application of the DHC. In this connection,
it is to be pointed out that the explicit expressions of $P_r^{(d)}, \, F^{(d)}$ and $ \bar F^{(d)}$
have been already given in Eq. (20) and we have the expression for the super-exterior derivative
as $\tilde d = dt\, \partial_t + d\theta \, \partial_{\theta} + d\bar\theta\, \partial_{\bar\theta}$.
Written in the explicit forms, the {\it quantum} dual gauge [i.e. (anti-)co-BRST] transformation (61)
implies the following expressions  for $P^{(d)}_{(r)}(t, \theta, \bar\theta), F^{(d)} (t, \theta, \bar\theta) $ 
and $\bar F^{(d)} (t, \theta, \bar\theta)$ in terms of the SUSP {\it dual} unitary operator 
and its Hermitian conjugate, namely;
\begin{eqnarray}
&&P_{r}^{(d)}(t, \theta, \bar\theta) = p_{r}(t)\,+\,i\,(\partial_{t}\,\tilde U)\,{\tilde U}^{\dagger}, \nonumber\\
&&F^{(d)}(t, \theta, \bar\theta) = i\,(\partial_{\theta}\,{\tilde U})\,{\tilde U}^{\dagger},\qquad\quad
\bar F^{(d)}(t, \theta, \bar\theta) = i\,(\partial_{\bar\theta}\,{\tilde U})\,{\tilde U}^{\dagger}.
\end{eqnarray} 
 The explicit substitution of $\tilde U$ and ${\tilde U}^{\dagger}$ from (59) 
into the above relationships yields exactly the same result as (20) for the expansions
of $P_{r}^{(d)}, F^{(d)}$ and $\bar F^{(d)}$.

In the above, we have constructed a 1-form $p_{(r)}^{(1)} = dt p_{(r)}(t)$ on the 1$D$ manifold
for the derivation of the (anti-)dual-BRST symmetries. This should
be contrasted with the 1-form $\lambda^{(1)} = dt \lambda(t)$ that was taken into account in the
context of the derivation of the (anti-)BRST symmetries [21]. We have done it because of the fact that there
is a duality  (i.e. $\lambda \to p_{r},\,C \to \bar C, \,\bar C \to C$) in the theory when we go from
$s_{(a)b} \to s_{(a)d}$. Thus, the super 1-form 
$\tilde \lambda^{(1)\;(h)}(t, \theta, \bar{\theta}) \to \tilde P_{(r)}^{(d)}(t, \theta, \bar{\theta})$
such that the appropriate super 1-form $P_{(r)}^{(1)}(t, \theta, \bar{\theta})$ is defined, for the
derivation of the (anti-)co-BRST symmetry transformations. Thus, now we have
$ P_{(r)}^{(1)}(t, \theta, \bar{\theta}) = dt\, P_{(r)}(t, \theta, \bar{\theta})\,+
\,d\theta\, F(t, \theta, \bar{\theta})\,+\,d\bar\theta \bar F(t, \theta, \bar{\theta})$.
We note that the above 1-form is derived from the definition of super 1-form
$ \tilde\lambda_{(r)}^{(1) (h)}(t, \theta, \bar{\theta}) = dt\, \Lambda^{(h)} (t, \theta, \bar{\theta})\,+
\,d\theta\,\bar F^{(h)} (t, \theta, \bar{\theta})\,+\,d\bar\theta F^{(h)} (t, \theta, \bar{\theta})$
that has been used for the derivation of the (anti-)BRST symmetries [21]. From 
relationship (61), it can be checked that $\tilde d \tilde P_{(r)}^{(1){(d)}} = 
d p_{r}(t) = 0 $ (where we have operated by $\tilde d$ from the left on 
$\tilde P_{(r)}^{(1){(d)}}$ and taken into account the fact that 
$\tilde d p_{(r)}^{(1)} = d p_{(r)}^{(1)} = 0 $  and 
$\tilde d \tilde U \wedge\, \tilde d \tilde U^{\dagger} = 0 $).
To be more precise, it can be checked that $\tilde d\, p_r^{(1)} = d\,p_r^{(1)}$
because $p_r^{(1)} = dt p_r(t)$ and $\partial_\theta \, p_r(t) = \partial_{\bar\theta} \, p_r(t) = 0 $.
The explicit form of $\tilde d\,\tilde U $ and $\tilde d\,\tilde U^\dagger $ are as follows:
\begin{eqnarray}
\tilde d\,\tilde U &=& dt\,\bigl[\theta\,(-\,i\,\dot C) +  \bar\theta\,(-\,i\,\dot {\bar C}) +
\theta\,\bar\theta \,\bigl(\dot B - \dot{\bar C}\,C - \bar C\,\dot C\bigr)\bigr] \nonumber\\
&+& d\theta\, \bigr[-\,i\, C + \bar\theta\, (B - \bar C\, C)\bigr] +
d{\bar\theta}\, \bigr[-\,i\, \bar C - \theta\, (B - \bar C\, C)\bigr],\nonumber\\
\tilde d\,\tilde U^\dagger &=& dt\,\bigl[\theta\,(i\,\dot C) +  \bar\theta\,(i\,\dot {\bar C}) +
\theta\,\bar\theta \,\bigl(-\,\dot B - \dot{\bar C}\,C - \bar C\,\dot C\bigr)\bigr] \nonumber\\
&+& d\theta\, \bigr[i\, C + \bar\theta\, (-\,B - \bar C\, C)\bigr] +
d{\bar\theta}\, \bigr[i\, \bar C - \theta\, (-\,B - \bar C\, C)\bigr].
\end{eqnarray}
The claim $\tilde d\,\tilde U \, \wedge \tilde d\,\tilde U^\dagger = 0 $ can be proven by 
collecting {\it all} the coefficients of $(dt\,\wedge \,d\theta), \, (dt\,\wedge \,d{\bar\theta}),\,
(d\theta\,\wedge \,d\theta), \,(d{\bar\theta}\,\wedge \,d{\bar\theta})$ and
$(d\theta\,\wedge \,d{\bar\theta})$ and showing that these are {\it exactly} zero. There is a
 simpler method to prove this statement by looking carefully at the exponential 
forms of $U $ and $ U^\dagger $ [cf. (60)]. We note
 that the {\it exponents} are the {\it same} modulo a sign factor. Therefore, the quantity
 $\tilde d\,\tilde U \wedge \tilde d\,\tilde U^\dagger$ would imply the wedge product 
between the same quantities (i.e. exponents). Since the exponents are bosonic in nature, their
wedge product would {\it always} be zero. Thus, we conclude 
that $\tilde d\,\tilde U \, \wedge \tilde d\,\tilde U^\dagger = 0 $ which implies that a
2-form ($\tilde d\, P_r^{(d)} = d\,p_r^{(1)} = 0$) cannot be defined on a 1$D$ manifold.
Hence, the r.h.s. of $\tilde d\,\tilde U \, \wedge \tilde d\,\tilde U^\dagger$ is zero.

We have to express the super expansion of $\Lambda^{(d)} (t, \theta, \bar\theta) $ in terms
of $U $ and $ U^\dagger $. In this connection, we observe that (anti-)dual BRST invariant
 quantity of interest is:
\begin{eqnarray}
s_{(a)d}\, \bigr[\,p_r - \dot\lambda\,\bigl] = 0. 
\end{eqnarray}
According to AVSF, we have the following equality 
$\Bigl($with $\lambda (t) \to \Lambda (t, \theta, \bar\theta)\Bigr)$
\begin{eqnarray}
 \tilde P_r^{(1)(d)}(t, \theta, \bar\theta) - \tilde d\, \Lambda(t, \theta, \bar\theta) = 
p_r^{(1)} (t) - d\,\lambda (t),
\end{eqnarray}
where the other symbols have been explained earlier. Taking the input from Eq. (61), 
we obtain the following:
\begin{eqnarray}
 p_r(t) + i\,(\tilde d \, \tilde U)\,{\tilde U}^\dagger
  -  \tilde d\,\Lambda(t, \theta, \bar\theta) = p_r^{(1)} (t) - \dot\lambda (t).
\end{eqnarray}
From the above relationship, it is very much evident that we obtain the following relations:
\begin{eqnarray}
&&\dot\Lambda = \dot\lambda + i\,(\partial_t \, \tilde U)\,{\tilde U}^\dagger, \qquad
\partial_\theta \,\Lambda = i\,(\partial_\theta \, \tilde U)\,{\tilde U}^\dagger,
\qquad \partial_{\bar\theta} \,\Lambda = i\,(\partial_{\bar\theta} \, \tilde U)\,{\tilde U}^\dagger.
\end{eqnarray}
It is very interesting to observe, from Eq. (63), that we have the relationships
$\partial_\theta \,\Lambda (t, \theta, \bar\theta)  = F^{(d)} (t, \theta, \bar\theta) $ 
and $\partial_{\bar\theta} \,\Lambda (t, \theta, \bar\theta) = {\bar F}^{(d)} (t, \theta, \bar\theta) $.
Thus, taking into account the expansions given in (63) and (67), we have derived the expansion
$\Lambda^{(d)} (t, \theta, \bar\theta)$ [cf. (20)] in terms of the SUSP {\it dual}
unitary operator and its Hermitian conjugate. Ultimately, we note that the (anti-)co-BRST symmetry invariance
of the quantities (i.e. $s_{(a)}d\,\bigl[r, \theta, p_\theta\bigr]$ = 0) can be translated 
into the generalizations: $r(t) \to R(t, \theta, \bar\theta) = 
r(t), \theta (t) \to \Theta (t, \theta, \bar\theta) = \theta(t)$ and $p_\theta (t)
\to P_\theta (t, \theta, \bar\theta) = p_\theta (t)$ which are trivial generalizations.

We concentrate now on the modified version of 2$D$ Proca theory as well as the anomalous
gauge theory and express the DHC and DGIRs in terms of the SUSP {\it dual} unitary 
operator and its Hermitian conjugate. In these theories, there is a duality in the sense
that the transformations: $A_\mu \to  A^{(d)}_\mu = -\;\varepsilon_{\mu\nu}\,A^\nu, \, C \to \bar C,
 \, \bar C \to C$ yield 
the (anti-)dual BRST symmetry transformations $s_{(a)d}$ from the (anti-)BRST symmetry 
transformations $s_{(a)b}$ for the gauge and (anti-)ghost fields [21]. Thus, we define the 
{\it dual} super 1-form connection, as an input for the derivation of the 
(anti-)co-BRST symmetries, as:
\begin{eqnarray}
\tilde A_{(d)}^{(1)}(x, \theta, \bar\theta) = d x^{\mu}\,\bigl[-\,\varepsilon_{\mu\nu}\,
{\cal B}^{\nu}_{(d)}(x, \theta, \bar\theta)\bigr] 
+ d\theta\,F^{(d)}(x, \theta, \bar\theta) + d\bar\theta\,\bar F^{(d)}(x, \theta, \bar\theta).
\end{eqnarray}
It is to be noted that we have derived ${\tilde A}^{(1)}_{(d)}(x, \theta, \bar\theta)$ from
the {\it usual} super 1-form ${\tilde A}^{(1)} = dx^{\mu}\,{\cal B}_{\mu}(x, \theta, \bar\theta)
+ d\theta \bar F(x, \theta, \bar\theta) + d\bar\theta F(x, \theta, \bar\theta) $ by the 
replacements: ${\cal B}_\mu \to -\, \varepsilon_{\mu\nu}\,{\cal B}^\nu, \, F \to \bar F$ 
and $\bar F \to F$ due to the presence of {\it duality} in our theory. In the context of
(anti-)co-BRST symmetries, it will be noted that the {\it usual} definition of the super
1-form (i.e. $\tilde A^{(1)}$) is taken into account in a subtle manner. Under the (anti-)dual 
BRST symmetry transformations, the above super 1-form transforms in the superspace as
\begin{eqnarray}
\tilde A_{(d)}^{(1)} (x, \theta, \bar\theta) = \tilde U (x, \theta, \bar\theta) \, \tilde A^{(1)} (x) \,
{\tilde U}^\dagger (x, \theta, \bar\theta) + i (\tilde d \tilde U)\,{\tilde U}^\dagger,
\end{eqnarray}
where, for the modified version of 2$D$ Proca theory, the form of SUSP dual unitary operator $\tilde U$
and its Hermitian conjugate ${\tilde U}^\dagger$ are:
\begin{eqnarray}
&&\tilde U\,(x, \theta, \bar{\theta})\,=\,exp\,[\theta\,(-\,i\,\bar{C})\,+
\,\bar{\theta}\,(-\,i\,C)\,+\,\theta\,\bar{\theta}\,\{-\, (E - m\,\tilde\phi\})],\nonumber\\
&&U^{\dagger}\,(x, \theta, \bar{\theta})\,=\,exp\,[\theta\,(i\,\bar{C})\,+
\,\bar{\theta}\,(i\,C)\,+\,\theta\,\bar{\theta}\,(E - m\,\tilde\phi)].
\end{eqnarray} 
The above can be explicitly written (in terms of coefficients of $\theta, \bar\theta$ and 
$\theta\,\bar\theta$) as:
\begin{eqnarray}
&&\tilde U\,(x, \theta, \bar{\theta})\,=\,1+\,\theta\,(-\,i\,\bar{C})\,+\,\bar{\theta}\,(-\,i\,C)\,
+\,\theta\,\bar{\theta}\,[-\,(E - m\,\tilde\phi)\,-\,\bar C\,{C}],\nonumber\\
&&\tilde U^{\dagger}\,(x, \theta, \bar{\theta})\,=\,1+\,\theta\,(i\,\bar{C})\,+\,\bar{\theta}\,(i\,C)\,
+\,\theta\,\bar{\theta}\,[(E - m\,\tilde\phi)\,-\,\bar{C}\,{C}].
\end{eqnarray}
The substitution of (72) into (70) yields the following 
\begin{eqnarray}
{\bar F}^{(d)} =  i\,(\partial_{\bar\theta} \, \tilde U)\,{\tilde U}^\dagger, 
\quad {F}^{(d)} =  i\,(\partial_{\theta} \, \tilde U)\,{\tilde U}^\dagger, 
\quad \varepsilon_{\mu\nu}\, {\cal B}^{\nu}(x, \theta, \bar\theta) = 
\varepsilon_{\mu\nu}\, {A}^{\nu} (x) - i\,(\partial_{\mu} \, \tilde U)\,{\tilde U}^\dagger, 
\end{eqnarray}
where we have equated the coefficients of $dx^\mu, d\theta$ and $d\bar\theta$ from the
l.h.s. and r.h.s. of (70). The last entry in the above equation leads to the following:
\begin{eqnarray}
{\cal B}_{\mu}^{(d)} &=& A_\mu (x) - i\,\varepsilon_{\mu\nu}\,(\partial^\nu \tilde U)\,
 {\tilde U}^\dagger\nonumber\\
&\equiv & A_\mu (x) + \theta (-\,\varepsilon_{\mu\nu}\, \partial^\nu C)
+ \bar\theta (-\,\varepsilon_{\mu\nu}\, \partial^\nu \bar C) 
+ \theta\,\bar\theta \bigl[ i\, \varepsilon_{\mu\nu} (E - m\,\tilde\phi)\bigr]\nonumber\\
&\equiv & A_\mu (x) + \theta (s_{ad}\, A_\mu) +  \bar\theta (s_{d}\, A_\mu) +
\theta\,\bar\theta \, (s_{d}\,s_{ad}\, A_\mu).
\end{eqnarray}
Similarly, we have the following super expansions in an explicit form: 
 \begin{eqnarray}
{F}^{(d)} &=&  i\,(\partial_{\theta} \, \tilde U)\,{\tilde U}^\dagger \, =\,
C(x) + \bar\theta\, (-\, \bigl[E - m\,\tilde\phi\bigr]) \,
\equiv \,  C(x) + \bar\theta \,(s_{d}\, C),\nonumber\\
{\bar F}^{(d)} &=&  i\,(\partial_{\bar\theta} \, \tilde U)\,{\tilde U}^\dagger \,=\,
\bar C(x) + \theta \,(\bigl[E - m\,\tilde\phi]\bigr)\,
\equiv \, \bar C(x) + \theta \,(s_{ad}\, \bar C).
\end{eqnarray}
Thus, we have derived the proper (anti-)co-BRST symmetry transformations for the basic
fields $A_\mu (x)\,C(x)$ and $\bar C(x)$ which are {\it common} for the modified versions of the
2$D$ Proca and anomalous gauge theories. In the latter case, however, we have to replace
$(E - m\,\tilde\phi)$  by ${\cal B}(x)$ in the definition of the SUSP {\it dual} unitary 
operator and its Hermitian conjugate.  We shall now focus on the derivation of proper
(anti-)co-BRST symmetries for the additional fields in {\it these} theories.

Let us express the superfield $\tilde \phi^{(d)} (x, \theta, \bar\theta)$ in the
language of the SUSP {\it dual} unitary operator and its Hermitian conjugate. We have
seen that $s_{(a)d}\,[E - m\,\tilde\phi] = 0$ due to the on-shell nilpotency
(i.e. $s_{(a)d}^2 \,C =  s_{(a)d}^2 \,\bar C =0$) in the theory because of the fact
that the (anti-)ghost fields $(\bar C)\;C$ obey the on-shell conditions: 
$(\Box + m^2)\;C = 0, \; (\Box + m^2)\;\bar C = 0$. Thus, the combination 
$[E - m\,\tilde\phi]$ is an (anti-)co-BRST invariant quantity which can be 
generalized onto the (2, 2)-dimensional supermanifold due to AVSF. This 
can be expressed in the language of superfields, derived after the application 
of the DHCs and DGIRs, as:
\begin{eqnarray}
\varepsilon^{\mu\nu}\,\partial_\mu\;{\cal B}_{\nu}^{(d)} (x, \theta, \bar\theta)
+ m\,\tilde\Phi (x, \theta, \bar\theta) = \varepsilon^{\mu\nu}\;\partial_\mu\; A_\nu (x)
+ m\;\tilde\phi (x).
\end{eqnarray}
Using the expression for ${\cal B}_{\mu}^{(d)} (x, \theta, \bar\theta)$ from (74),
we have
\begin{eqnarray}
\varepsilon^{\mu\nu}\,\partial_\mu\;\bigl[A_\nu (x) - i\,\varepsilon_{\nu\lambda}\,
(\partial^\lambda \tilde U)\, {\tilde U}^\dagger\bigr]
+ m\,\tilde\Phi (x, \theta, \bar\theta) = \varepsilon^{\mu\nu}\;\partial_\mu\; A_\nu (x)
+ m\;\tilde\phi (x).
\end{eqnarray}
This relation, finally, leads to the following expression for $\tilde\Phi^{(d)} (x, \theta, \bar\theta)$
in terms of the SUSP dual unitary operators (i.e.  $\tilde U$ and ${\tilde U}^\dagger$), namely;
\begin{eqnarray}
\tilde\Phi^{(d)} (x, \theta, \bar\theta) = \tilde\phi (x) 
+ \frac{i}{m}\,(\partial_\mu\,\tilde U)\,(\partial^\mu\;{\tilde U}^\dagger)
+ \frac{i}{m}\,(\Box\, \tilde U)\;{\tilde U}^\dagger.
\end{eqnarray}
It is very interesting to check that the r.h.s. yields the expansions (34) when we use 
the on-shell conditions: $(\Box + m^2)\;C = 0, \; (\Box + m^2)\;\bar C = 0$ and 
$(\Box + m^2)\;{\cal B} = 0$ . It is important to point out that the contributions, from the
{\it second} term of (78), cancel out with the extra piece that emerges from the {\it last} term 
on the r.h.s. of (78). Thus, we have expressed {\it all} the non-trivial nilpotent and
anticommuting (anti-)co-BRST symmetry transformations of the 2$D$ Proca theory in terms of the
SUSP dual unitary operators $\tilde U$ and ${\tilde U}^\dagger$.

A close look and careful observations of the equations (34) and (36) demonstrate that
the expansions are very similar and they differ only by a factor of $m$. Thus, it is
very elementary to note that the expansions (36) can be expressed in terms of the SUSP operators
(i.e. SUSP unitary operator $\tilde U$ and its Hermitian conjugate ${\tilde U}^\dagger$)
as follows:
\begin{eqnarray}
\Phi^{(d)} (x, \theta, \bar\theta) = \phi (x) +
i\,(\partial_\mu\,\tilde U)\,(\partial^\mu\;{\tilde U}^\dagger) +
i\, (\Box\, \tilde U)\;{\tilde U}^\dagger.
\end{eqnarray}
The substitution of the expressions for $\tilde U$ and ${\tilde U}^\dagger$
from (71) and (72) (with the replacement $(E - m\;\tilde\phi) \to {\cal B}$), we obtain the 
r.h.s. of the expansion (36) from the r.h.s. of the above relationship. We concentrate now on the
alternative to the expansion $\Sigma^{(d)} (x, \theta, \bar\theta)$ (cf. Eq. (40)) in the language of the
SUSP unitary operator  $\tilde U$  and its Hermitian conjugate ${\tilde U}^\dagger$. This can 
be derived  from the restrictions  (due to $s_{(a)d}\; [(a - 1) \; \sigma (x) - \phi (x) = 0]$) 
on the superfields, due to the basic tenets of AVSF, as: 
\begin{eqnarray}
(a - 1)\;\Sigma (x, \theta, \bar\theta) - \Phi^{(d)} (x, \theta, \bar\theta) =
(a - 1) \; \sigma (x) - \phi (x).
\end{eqnarray}
The substitution of (79) into the above equation yields the following
\begin{eqnarray}
\Sigma^{(d)} (x, \theta, \bar\theta) = \sigma (x) + 
\frac{i}{(a - 1)}\;(\partial_\mu\,\tilde U)\,(\partial^\mu\;{\tilde U}^\dagger)
+ \frac{i}{(a - 1)}\; (\Box\, \tilde U)\;{\tilde U}^\dagger.
\end{eqnarray}
Thus, we have obtained {\it all} the non-trivial (anti-)co-BRST symmetry
transformations for the modified version of 2$D$ anomalous gauge theory in the
terminology of SUSP dual unitary operator $\tilde U$  and its Hermitian conjugate
${\tilde U}^\dagger$. In other words, we conclude that the precise derivations of 
$\tilde U$  and  ${\tilde U}^\dagger$ provide the alternatives to the DHCs and DGIRs that
are exploited within the framework of AVSF for the derivation of the 
(anti-)co-BRST symmetry transformations. Thus, the precise forms of $\tilde U$ 
 and  ${\tilde U}^\dagger$ are {\it physically} important.

Finally, we focus on the alternative to the DHC and DGIRs in the context of 
2$D$ self-dual bosonic field theory. Here the SUSP dual unitary operators
$\tilde U$  and  ${\tilde U}^\dagger$ would be {\it exactly} same as in Eq. (59)
with some replacements in view of the (anti-)co-BRST symmetry transformations (2)
{\it vis-{\`a}-vis} Eq. (9). Thus, we have now 
\begin{eqnarray}
&&\tilde U\,(x, \theta, \bar{\theta})\,=\,exp\,\Bigl[\theta\,(-\,i\,\bar{C})\,+
\,\bar{\theta}\,(-\,i\,C)\,+\,\theta\,\bar{\theta}\,
\bigl\{-\; \frac{1}{2}(\dot \phi\,-\,\dot v\,+\,v'\,-\,\phi')\bigr\}\Bigr],\nonumber\\
&&\tilde U^{\dagger}\,(x, \theta, \bar{\theta})\,=\,exp\,\Bigl[\theta\,(i\,\bar{C})\,+
\,\bar{\theta}\,(i\,C)\,+\,\theta\,\bar{\theta}\,
\bigl\{ \frac{1}{2}(\dot \phi\,-\,\dot v\,+\,v'\,-\,\phi')\bigr\}\Bigr].
\end{eqnarray} 
We define the {\it dual} super 1-form as follows:
\begin{eqnarray}
\Phi^{(1)} (x, \theta, \bar{\theta}) = dt\; (2\; \Phi^{(d)} (x, \theta, \bar{\theta}) 
+ d\theta \; F^{(d)} (x, \theta, \bar{\theta}) + d\bar\theta\; {\bar F^{(d)}} (x, \theta, \bar{\theta}),
\end{eqnarray} 
where the expansions for $\Phi^{(d)} (x, \theta, \bar{\theta}), F^{(d)} (x, \theta, \bar{\theta})$
and ${\bar F^{(d)}} (x, \theta, \bar{\theta})$ have to be determined in terms of 
$\tilde U$  and  ${\tilde U}^\dagger$  listed in (82). We point out that a factor of {\it two} has been taken into
account in (83) because of the observation that ($s_{d}\;\phi = \dot{\bar C}/2,\;
s_{ad}\;\phi = \dot{C}/2$) [cf. (9)]. The transformations of the super 1-form (83) in the superspace is 
\begin{eqnarray}
\Phi^{(1)\;(d)} (x, \theta, \bar{\theta}) = \tilde U \; \phi^{(1)}(x) \;{\tilde U}^\dagger 
+ i\, (\tilde d\;\tilde U)\;{\tilde U}^\dagger, 
\end{eqnarray} 
where $\phi^{(1)} (x) = dt\; [2\; \phi (x)]$ is a 1-form on the 1$D$ sub-manifold of 
the general 2$D$ ordinary spacetime manifold. The above equation, taking into account
the definition (83), is as follows in the component form
\begin{eqnarray}
&&2\,\Phi^{(d)} (x, \theta, \bar{\theta}) =  2\;\phi (x) + i\,(\partial_t \;\tilde U){\tilde U}^\dagger, 
\nonumber\\
&& F^{(d)} (x, \theta, \bar\theta) = i\,(\partial_\theta \;\tilde U){\tilde U}^\dagger,
\qquad {\bar F}^{(d)} (x, \theta, \bar\theta) = i\,(\partial_{\bar\theta} \;\tilde U){\tilde U}^\dagger,
\end{eqnarray} 
where we have taken into account  the comparison of the coefficients of $dx^\mu, d\theta$ and 
$d\bar\theta$ from r.h.s. and l.h.s. The substitution of the explicit form of $\tilde U$ 
and  ${\tilde U}^\dagger$ from (82) leads to the following expansions from the superfields
(cf. Eq. (55)): 
\begin{eqnarray}
&&\Phi^{(d)} (x, \theta, \bar{\theta}) =  \phi (x) + \theta \; \Bigl(\frac{\dot C}{2}\Bigr) 
+ \bar\theta \Bigl(\frac{\dot {\bar C}}{2}\Bigr) 
+ \theta\;\bar\theta \Bigl(-\; \frac{i}{4}\,\frac{\partial}
{\partial t}\;[\dot \phi\,-\,\dot v\,+\,v'\,-\,\phi']\Bigr),\nonumber\\ 
&& F^{(d)} (x, \theta, \bar\theta) = C(x) + \bar\theta \; 
\Bigl(-\; \frac{i}{2}\,[\dot \phi\,-\,\dot v\,+\,v'\,-\,\phi']\Bigr),\nonumber\\ 
&&{\bar F}^{(d)} (x, \theta, \bar\theta) = {\bar C}(x) + \theta \;
 \Bigl(\frac{i}{2}\,[\dot \phi\,-\,\dot v\,+\,v'\,-\,\phi']\Bigr). 
\end{eqnarray}  
A close look at the expansions shows that we have already derived the (anti-)co-BRST 
symmetry transformations $s_{(a)d}$ for the fields $\phi (x), C(x), \bar C(x)$. 
We note that $s_{(a)d}\; [\phi - v] = 0$. This observation implies immediately,
due to the basic tenets of AVSF, that we have the following expansion of the superfield
corresponding to the WZ-field $v(x)$, namely;
\begin{eqnarray}
v(x) \to V^{(d)} (x, \theta, \bar{\theta}) &=& v(x) + \theta \; \Bigl(\frac{\dot C}{2}\Bigr) 
+ \bar\theta \Bigl(\frac{\dot {\bar C}}{2}\Bigr) 
+ \theta\;\bar\theta \Bigl(-\; \frac{i}{4}\,\frac{\partial}{\partial t}
[\dot \phi\,-\,\dot v\,+\,v'\,-\,\phi']\Bigr),\nonumber\\ 
&\equiv & v(x) + \theta \Bigl(s_{ad}\, v(x)\Bigr) + \bar\theta \Bigl(s_d\, v(x)\Bigr)
 + \theta\,\bar\theta\, \Bigl(s_d\,s_{ad}\,v(x)\Bigr),
\end{eqnarray}   
where the (anti-)co-BRST symmetry transformations $s_{(a)d}$ are listed in (9). 
The above equation (87) can also be written in terms of  $\tilde U$  and  ${\tilde U}^\dagger$ as: 
\begin{eqnarray}
2\,V^{(d)} (x, \theta, \bar{\theta}) =  2\;v (x) + i\,(\partial_t \;\tilde U){\tilde U}^\dagger. 
\end{eqnarray}   
This is due to the fact that a super 1-form can be written exactly like (83) in terms of
$V^{(d)} (x, \theta, \bar{\theta}), F^{(d)} (x, \theta, \bar\theta)$ and 
${\bar F}^{(d)} (x, \theta, \bar\theta)$. It goes without saying that we  can repeat the
above exercise to obtain the superspace transformation like (87) and (88).

We observe that $s_{(a)d}\;\bigl[\dot \lambda - 2\;\phi\bigr] = s_{(a)d}\;\bigl[\dot \lambda - 2\;v\bigr] = 0$. 
Thus, we have the following restrictions (due to these invariances) on the superfields,
 defined on (2, 2)-dimensional supermanifold, according to
 to basic tenets of AVSF, namely;
\begin{eqnarray} 
&&\dot \Lambda^{(d)} (x, \theta, \bar{\theta}) - 2\; \Phi^{(d)} (x, \theta, \bar{\theta}) =
 \dot \lambda(x) - 2\;\phi(x),\nonumber\\
&&\dot \Lambda^{(d)} (x, \theta, \bar{\theta}) - 2\; V^{(d)} (x, \theta, \bar{\theta}) = 
\dot \lambda(x) - 2\;v(x),
\end{eqnarray}
which implies that the superfield $\Lambda^{(d)} (x, \theta, \bar{\theta})$ can be expressed 
(from both the above relationships) in terms of the
SUSP dual unitary operators $\tilde U$  and  ${\tilde U}^\dagger$ as: 
\begin{eqnarray}
\dot\Lambda^{(d)} (x, \theta, \bar{\theta}) = \dot \lambda (x) + i\,(\partial_t \;\tilde U){\tilde U}^\dagger.
\end{eqnarray}
The above expression finally leads to: 
\begin{eqnarray}
\Lambda^{(d)} (x, \theta, \bar{\theta}) &=&  \lambda (x) + + \theta \;(C) + \bar\theta (\bar C) 
+ \theta\;\bar\theta \Bigl(-\; \frac{i}{2}\,[\dot \phi\,-\,\dot v\,+\,v'\,-\,\phi']\Bigr),\nonumber\\ 
&\equiv & \lambda(x) + \theta \bigl(s_{ad}\, \lambda(x)\bigr) + \bar\theta \bigl(s_d\, \lambda(x)\bigr) 
+ \theta\,\bar\theta\, \bigl(s_d\,s_{ad}\,\lambda(x)\bigr).
\end{eqnarray}
Thus, we have provided the alternatives to the DHC and DGIRs used in Sec. 3, in the language of 
$\tilde U$  and  ${\tilde U}^\dagger$ and obtained {\it all} the non-trivial (anti-)co-BRST symmetries of the
2$D$ self-dual bosonic field theory. We conclude this section with the remarks that SUSP dual unitary 
operators $\tilde U$  and  ${\tilde U}^\dagger$ provide the alternatives to the DHC and DGIRs within
the framework of AVSF where the explicit group structure is maintained.\\

\section{Conclusions}

For the Abelian 1-form $U(1)$ gauge theories, it is important to 
have explicit existence and appearance  of the {\it group structure} in any kind of computation.
The SUSP {\it dual} unitary operator and its Hermitian conjugate {\it do} exactly the same job in
our present endeavor and, that is why, their derivation is important. In our earlier works [26,27],
we have explicitly derived the {\it exact} form of the SUSP {\it unitary} operator and its Hermitian conjugate
for the cases of the interacting (i) 4$D$ Abelian $U(1)$ gauge theory with the Dirac and complex 
scalar fields, and (ii) 4D non-Abelian $SU(N)$ gauge theory with Dirac fields, in the context of
nilpotent (anti-)BRST symmetries. 
The {\it universal} nature of the SUSP unitary operator and its Hermitian conjugate has also been established 
in our recently published work [21] for the case of the 1$D$ and 2$D$ {\it Abelian} $U(1)$ gauge theories. 
In fact, we have been able to derive the {\it dual unitary} operator and its Hermitian conjugate from the
{\it above universal} unitary operator by exploiting the virtues of the {\it duality} symmetry 
in our theory where $C \to \bar C, \bar C \to C$ and $A_\mu \to -\, \varepsilon_{\mu\nu} \, A^\nu$.
As it turns out, we observe that the mathematical form of the SUSP {\it dual unitary} operator and its
Hermitian conjugate is {\it universal} in exactly the same way as the SUSP {\it unitary}
operator and its Hermitian conjugate are (see, e.g. [21]).

We would like to dwell a bit on the {\it duality} aspects of our statement. In the case of 2$D$ 
Abelian 1-form gauge theory, it can be seen that the self-duality condition:
$\ast \, A^{(1)} = \ast (dx^\mu \,A_\mu) = \varepsilon^{\mu\nu}\,dx_\nu\, A_\mu
\equiv dx^\mu\,(-\,\varepsilon_{\mu\nu}\,A^\nu) \equiv dx^\mu\, A_\mu^{(d)}$
where $A_\mu^{(d)} = -\,\varepsilon_{\mu\nu}\,A^\nu$ is the {\it dual} 1-form
potential corresponding to the Abelian 1-form potential $A_\mu$. Furthermore, we observe that
when we go from the (anti-)BRST symmetries (particularly in the ghost sector of our
theory), there is a transformation from $C \to \bar C$ and $\bar C \to C$. Thus, for a 2$D$ 1-form
theory, the transformations $ A_\mu \to -\,\varepsilon_{\mu\nu}\,A^\nu$,
$C \to \bar C$, $\bar C \to C$ are the {\it duality} transformations which have been exploited
in the definition of super 1-forms (cf. (69), (83)). However, in the case of 1$D$ Abelian 1-form theory
(i.e. a rigid rotor), we observe that there is a duality: $\lambda \to p_r, C \to \bar C, \bar C \to C$.
This observation has been exploited in the statements that have followed equations (61) and (63) in the definition
of $P^{(1)}_{(d)} (x, \theta, \bar\theta)$. Similar kind of arguments have been exploited
in the case of 2$D$ self-dual field theory where we have expressed the DHC and DGIRs 
in the language of  $\tilde U$  and  ${\tilde U}^\dagger$.

In our present endeavor, we have applied the AVSF to derive the (anti-)co-BRST symmetry transformations for
a {\it new} model in 2$D$. This model is nothing but the modified version of the 2D anomalous 
gauge theory which has already been proven to provide a  tractable model for the Hodge theory [19].
Thus, it is a {\it novel} result in our present endeavor. The precise derivation of the 
(anti-)co-BRST symmetries establishes the sanctity and correctness of the working-rule
that has been laid down  for the Hodge duality ($\star$) operation on the (1, 2) and 
(2, 2)-dimensional supermanifolds [22]. 
Thus, we conclude that the AVSF is a powerful theoretical technique that can be applied to 
interesting physical systems and one can derive the appropriate form of the BRST-type symmetries. 
The key concepts (that play important roles in the application of the AVSF) are the DHC and DGIRs.
One of the key observations of our present endeavor is the fact that the geometrical meaning of the 
(anti-)co-BRST symmetries, in the language of the translational generators 
($\partial_\theta, \, \partial_{\bar\theta}$) along the 
Grassmannian directions of the appropriately  chosen supermanifold, remains the same when we
exploit the theoretical strength of the DHC and DGIRs.

We would like to lay emphasis on the fact that the models of the Abelian 1-form gauge 
theories in 1$D$ and 2$D$ (that have been considered in our present endeavor) are 
{\it interesting} because these models provide the tractable physical examples of Hodge 
theory within the framework of BRST formalism [16-20]. Such models are mathematically 
as well as physically very rich because there are many continuous symmetries in the theory
which enable these theories to be quantized {\it without} the definition of the canonical conjugate
 momenta corresponding to the fields of these theories [29-32]. In the context of gauge 
theories, it has been shown, in our earlier works [29-32], that there exist {\it six} 
continuous {\it internal} symmetries {\it for such theories} which are so powerful that they 
lead to the canonical quantization of these theories at the level of creation and annihilation 
operators. The above symmetries have also played very important roles in the proof of 2$D$ (non-)Abelian
1-form gauge theories (without any interaction with matter fields) to be a {\it new} class [33] of topological 
field theories (TFTs) that capture a few key aspects of the Witten-type TFTs and some salient features of
the Schwartz-type TFTs.

We have succeeded in obtaining {\it universal} SUSP unitary operator and its Hermitian conjugate that are  
primarily connected with the (anti-)BRST symmetries in the cases of 4$D$ interacting
Abelian 1-form gauge theories with Dirac fields, 2$D$ and 1$D$ Abelian gauge theories. In our present 
endeavor, we have obtained the SUSP {\it dual} unitary operator and its Hermitian conjugate
in the cases of 2$D$ and 1$D$ Abelian 1-form gauge theories that are connected with the (anti-)co-BRST
symmetry transformations. One of the immediate goal for us is to extend our work to the 2$D$ non-Abelian
1-form gauge theory (without any interaction with matter fields) so that we could derive the 
SUSP unitary operator and its Hermitian conjugate as well as the SUSP {\it dual} unitary
operator and its Hermitian conjugate. This is essential because we have already shown that {\it this}
2$D$ non-Abelian model is an example of the Hodge theory where the (anti-)BRST and (anti-)co-BRST symmetries 
exist along with {\it other} internal symmetries. We have already made some progress in this direction and our
results would be reported in our future publication [34].\\

\vskip 0.7cm

\noindent
{\bf Competing Interests}\\

\noindent
The authors declare that they have no competing interests.\\

\vskip 0.7cm

\noindent
{\bf Acknowledgements} \\

\noindent 
One of us (TB) would like to gratefully acknowledge the financial 
support from CSIR, Govt. of India, New Delhi, under its SRF-scheme. Another author (NS) is thankful to the 
BHU-fellowship for financial support. The present investigation has been carried out under
the above financial supports.\\

\end{document}